\documentclass[12pt]{article}

\usepackage[dvips]{graphicx}
\usepackage{amssymb}
\usepackage{amsmath}
\usepackage{epsfig}

\usepackage{epsf}
\usepackage{graphicx,epsfig}
\usepackage{amsfonts}
\usepackage{amssymb}

\def\nn{\nonumber}

\def\del{\partial}

\def\CV{{\cal V}}

\makeatletter
\renewcommand\section{\@startsection {section}{1}{\z@}%
                                 {-3.5ex \@plus -1ex \@minus -.2ex}
                                   {2.3ex \@plus.2ex}%
                                   {\normalfont\large\bfseries}}
\renewcommand\subsection{\@startsection{subsection}{2}{\z@}%
                                   {-3.25ex\@plus -1ex \@minus -.2ex}%
                                     {1.5ex \@plus .2ex}%
                                     {\normalfont\bfseries}}
\renewcommand\subsubsection{\@startsection{subsubsection}{3}{\z@}%
                                   {-3.25ex\@plus -1ex \@minus -.2ex}%
                                     {1.5ex \@plus .2ex}%
                                     {\normalfont\itshape}}
\makeatother

\newcommand{\be}{\begin{equation}}
\newcommand{\ee}{\end{equation}}
\newcommand{\bea}{\begin{eqnarray}}
\newcommand{\eea}{\end{eqnarray}}
\newcommand{\barr}{\begin{array}}
\newcommand{\earr}{\end{array}}

\def\beq{\begin{equation}}
\def\eeq{\end{equation}}
\def\be{\begin{equation}}
\def\ee{\end{equation}}
\def\bea{\begin{eqnarray}}
\def\eea{\end{eqnarray}}
\def\d{{\rm d}}

\DeclareRobustCommand{\SkipTocEntry}[4]{}

\begin{document}

\begin{titlepage}

\setcounter{page}{1} \baselineskip=15.5pt \thispagestyle{empty}

\begin{flushright}
\end{flushright}
\vfil

\begin{center}
{\LARGE Monodromy in the CMB: \\[1.ex]  Gravity Waves and String Inflation}

\end{center}
\bigskip\

\begin{center}
{\large Eva Silverstein and Alexander Westphal}
\end{center}

\begin{center}
\textit{SLAC and Department of Physics, Stanford University, Stanford CA 94305}
\end{center} \vfil

\noindent  We present a simple mechanism for obtaining large-field inflation, and hence a gravitational wave
signature, from string theory compactified on twisted tori.  For Nil manifolds, we obtain a leading inflationary
potential proportional to $\phi^{2/3}$ in terms of the canonically normalized field $\phi$, yielding predictions
for the tilt of the power spectrum and the tensor-to-scalar ratio, $n_s\approx 0.98$ and $r\approx 0.04$ with 60
e-foldings of inflation; we note also the possibility of a variant with a candidate inflaton potential
proportional to $\phi^{2/5}$. The basic mechanism involved in extending the field range -- monodromy in D-branes
as they move in circles on the manifold -- arises in a more general class of compactifications, though our
methods for controlling the corrections to the slow-roll parameters require additional symmetries.

\vfil
\begin{flushleft}
March 20, 2008
\end{flushleft}

\end{titlepage}

\newpage
\tableofcontents
\newpage

\section{Introduction and Motivation}
\label{sec:intro}

It is of significant interest to develop inflationary mechanisms and models within string theory, as a UV
complete theory of gravity. One reason for this is that inflationary models are UV sensitive:   generic
dimension six Planck-suppressed operators contribute order 1 corrections to the (generalized) slow roll
parameters such as $\epsilon = -\frac{\dot{H}}{H^2}$.  These must be at most of order $10^{-2}$ so as to obtain
and sustain inflation -- nearly constant $H=\dot a/a$ -- for sufficiently long to solve the standard
cosmological problems by increasing the scale $a(t)$ of the universe by a factor of $e^{60}$ \cite{Inflation}.
Other, related motivations for studying inflation in string theory include its role in discovering new
mechanisms for inflation, and the potential for correlating (or anti-correlating) specific classes of
field-theoretic inflationary models and signatures with classes of string compactifications.\footnote{For
reviews of many works representing significant recent progress along these lines, see \cite{reviews}.}

In this paper, we propose a new class of string inflation models arising in compactifications on manifolds with
metric flux such as Nil manifolds which contain tori twisted over circles \cite{SS,ESdS}. A simple geometric
feature of these spaces -- monodromies in the one-cycles of the tori as one moves around the
circle\footnote{This effect was discovered in a broad class of backgrounds -- including non-geometric ones -- in
\cite{Albionmonodromy}.  It would be very interesting to study our mechanism in this larger setting.} -- leads
to large field ranges for the collective coordinates for wrapped D4-branes. This, combined with a detailed
analysis of the effective action and the dynamical constraints on the mutual consistency of inflation and moduli
stabilization, leads to a mechanism for large-field inflation with a $\phi^{2/3}$ potential for the canonically
normalized inflaton field $\phi$ (a variant of one of the earliest field-theoretic proposals for inflation
\cite{Linde:1983gd}).  As in \cite{Linde:1983gd}, we find that symmetries are very useful for helping to control
the inflaton trajectory.

Because the suppression of the generalized slow-roll parameters $\epsilon$ and $\eta$ must hold for every point
on the inflaton trajectory, it has a priori appeared particularly difficult to formulate large-field inflation,
in which the canonically normalized inflaton field $\phi$ rolls over a distance in field space large compared to
the four-dimensional Planck mass scale: $\Delta\phi\gg M_{\rm P}$.  These large-field models are especially
interesting from the observational point of view, because of the Lyth bound \cite{Lyth} showing that for any
single-field model of inflation, observable tensor modes in the CMB require a super-Planckian field range.
Upcoming CMB observations are projected to detect or constrain the tensor to scalar ratio $r$ at the level
$r\gtrsim 0.01$ (see for example \cite{Efstathiou:2006ak,Bmodeobs}), while existing and upcoming satellite
experiments also significantly constrain the tilt of the spectrum \cite{Observations}.

There are two types of conditions which must be satisfied in order to obtain large-field inflation, which we
will refer to as {\it geometrical} and {\it dynamical} criteria, respectively. First, the approximate moduli
space of the scalar inflaton must extend over a distance greater than $M_{\rm P}$.  Secondly the inflaton
dynamics in this region $\Delta\phi\gg M_{\rm P}$ must be consistent with the stabilization of the underlying
string compactification and with the smallness of the slow-roll parameters. In some interesting cases, even
obtaining the possibility, geometrically, of a large field range proved impossible \cite{FieldRange}. There are
other interesting mechanisms which do not exhibit such a geometric limitation -- such as \cite{Nflation}, a
theory of assisted inflation with multiple axions,\footnote{Axions can play an important role in small-field
string theoretic inflation as well; see \cite{Roulette}\ for an interesting recent example based partly on
\cite{Racetrack}.} and \cite{BLS}, brane inflation with multiple or wrapped branes rather than single D3-branes.
In many of these cases, the dynamical self-consistency criteria happened to become important parametrically
right at $\Delta\phi=M_{\rm P}$, leading to some speculation that a no-go theorem for gravity waves from string
inflation might hold.

Controlling inflation in string theory is a somewhat laborious process, as it requires stabilizing the
moduli\footnote{For reviews of various aspects as well as references of this subject, see e.g.
\cite{Fluxreviews}.}. This is because one must satisfy the generalized slow roll conditions in every direction
in field space in order to obtain inflation. Because moduli stabilization itself is somewhat complicated, and
outlined so far only for special corners of the theory where control is most easily available, our knowledge of
the space of possible top-down inflationary mechanisms and models is limited.

Recently a new class of de Sitter models in string theory was proposed in \cite{ESdS}\ in the context of
ten-dimensional type IIA string theory compactified on Nil (or Sol) manifolds with orientifold 6-planes,
fivebranes, and fluxes (complementing other work \cite{IIAnogo}\ showing that Calabi-Yau manifolds with a subset
of allowed ingredients can be cleanly excluded). In this class of string compactifications, we find that
inflation arises in a relatively straightforward way from the motion of a D4-brane wrapped on a one-cycle of the
compactification -- a brane which experiences a {\it monodromy}, not coming back to itself as it moves around a
circle in the manifold. The resulting geometrically large field range yields a candidate large-field model, in
which the brane executes multiple motions around the compact manifold during inflation.  The corresponding
inflaton potential descending from the D4-brane action turns out to be a fractional power, $V_{\cal R}(\phi)\sim
\mu^{10/3}\phi^{2/3}$ for the appropriate regime in $\phi$ in the simplest case which we analyze in
detail.\footnote{A slight variant of this yields a candidate $\phi^{2/5}$ potential which we will also discuss;
it would be interesting to analyze systematically the range of possibilities arising from branes with
monodromies in more general backgrounds with curvature and flux.} As far as dynamical criteria go, this
potential remains subdominant to the moduli potential over the requisite super-Planckian range of $\phi$, while
a related $m^2\phi^2$ region of the potential is cleanly excluded on these grounds.  We find further that the
problem of suppressing the slow-roll parameters $\epsilon$ and $\eta\sim \ddot\phi/(H\dot\phi)$ in this
background is alleviated by its symmetries, and by the fact that the brane circles the same manifold multiple
times during inflation. By considering moduli-fixing ingredients which are all extensive in the direction of the
motion of the brane,
we find that it is possible to avoid order 1 contributions to $\epsilon$ and $\eta$ which would otherwise be
there, and on this basis we argue for the existence of models realizing our mechanism.  Given that, the
(fractional) power law potential yields predictions for the tilt $n_s-1$ of the spectrum of density
perturbations, and the tensor to scalar ratio $r$, which we review in our final section.

\section{Towards D4-brane Inflation on Twisted Tori}
\label{sec:dynamics}

\subsection{Compactification Manifold, Monodromies, and Geometric Field Range}

The simplest example of a Nil manifold suffices to exhibit our mechanism for extending the field range.  A Nil
3-manifold is obtained by compactifying the nil geometry
\begin{eqnarray} \label{nilgeom} ds^2_{Nil} &=& L_{u_1}^2du_1^2+L_{u_2}^2du_2^2 + L_x^2\left(dx+\frac{M}{2}[u_1 d u_2-u_2 d
u_1]\right)^2 \nonumber \\
&=& L_{u_1}^2du_1^2+L_{u_2}^2du_2^2 + L_x^2\left(dx'+{M}u_1 d u_2\right)^2
\end{eqnarray}
(where $x'= x-\frac{M}{2}u_1u_2$) by a discrete subgroup of the isometry group
\begin{eqnarray} \label{translations}
 t_x: ~~ (x,u_1,u_2) &\to & (x+1, u_1,u_2)~, \nonumber \\  t_{u_1}: ~~(x,u_1,u_2) &\to &
 (x-\frac{M}{2}u_2, u_1+1,u_2)~, \nonumber \\ t_{u_2}:
~~ (x,u_1,u_2) &\to & (x+\frac{M}{2}u_1,u_1,u_2+1) \end{eqnarray}

This Nil 3-manifold ${\cal N}_3$ can be described as follows.  For each $u_1$, there is a torus in the $u_2$ and
$x'\equiv x-\frac{M}{2}u_1u_2$ directions. Moving along the $u_1$ direction, the complex structure $\tau$ of
this torus goes from $\tau\to \tau +M$ as $u_1\to u_1+1$. The projection by $t_{u_1}$ identifies these
equivalent tori.\footnote{The directions $u_1$ and $u_2$ are on the same footing; similar statements apply with
the two interchanged and with $x'$ replaced by $x''\equiv x+\frac{M}{2}u_1u_2$.}

At all values $u_1=j/M$ for integer $j$, the two-torus in the $u_2-x'$ directions is equivalent to a rectangular
torus
\be ds^2_{rect}\equiv L_x^2 dy_1^2+L_{u_2}^2dy_2^2~, ~~~~~ (y_1,y_2)\equiv (y_1+n_1,y_2+n_2) \ee
(since $\tau\to\tau+1$ as $j\to j+1$).  These coordinates $y_1$ and $y_2$ are related to $x'$ and $u_2$ by an
$SL(2,Z)$ transformation. The 1-cycle traced out by $u_2=\lambda,\lambda\in (0,1)$  becomes a cycle
$(y_1,y_2)=(M\lambda,\lambda)$ as $u_1\to u_1+1$.  Applied to wrapped branes, this monodromy will yield a simple
mechanism for obtaining a large field range, as follows.

\begin{figure}[ht]
\begin{center}
\includegraphics[width=15cm]{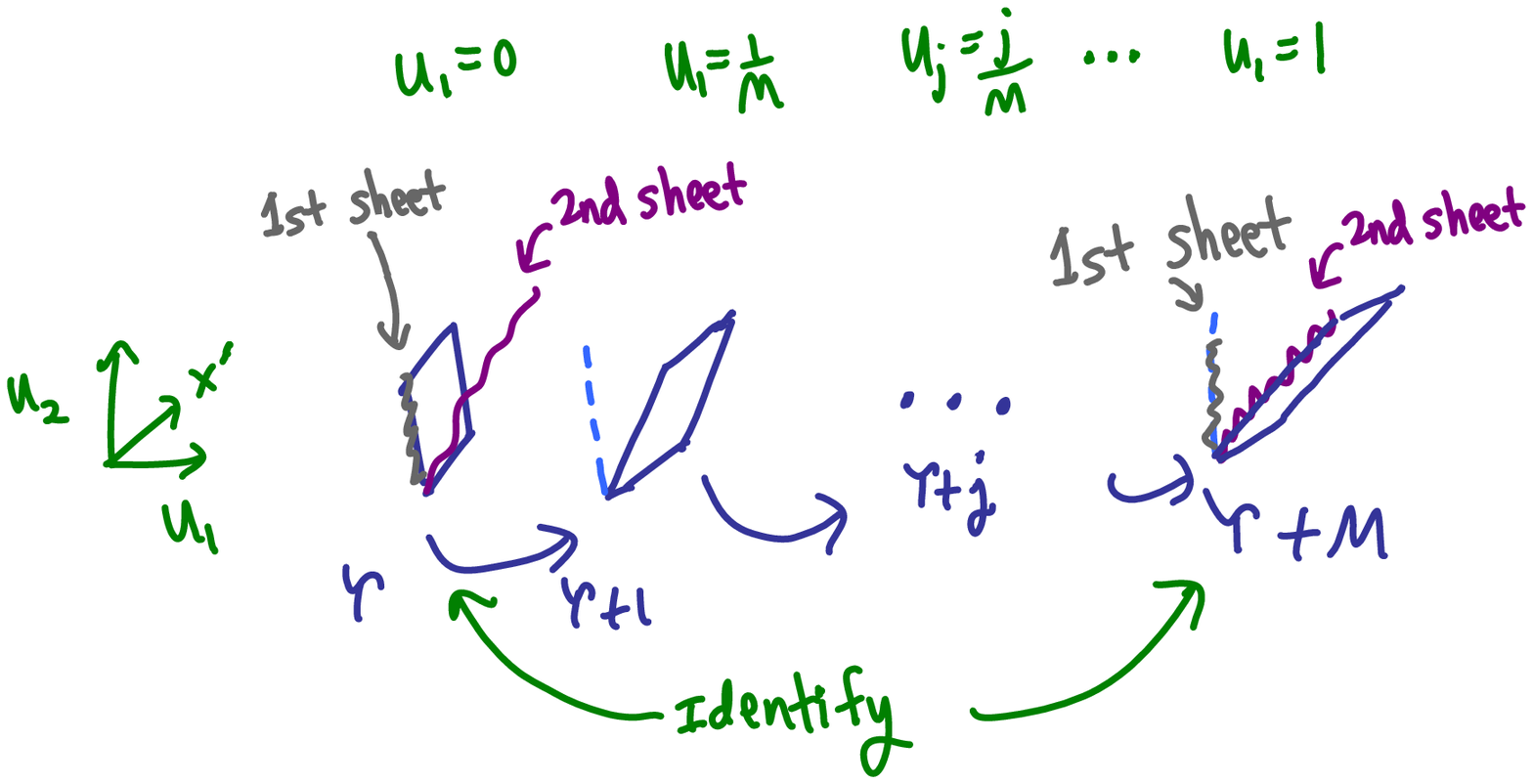}
\end{center}
\refstepcounter{figure}\label{fig:monodromy}

\vspace*{-.2cm} {\bf Figure~\ref{fig:monodromy}:} Monodromy of branes on a compact Nil 3-manifold.  Start with a
minimal-length D4-brane wrapped in the $u_2$ direction (indicated in grey). Moving it away from $u_1=0$
increases its mass since the cycle it wraps becomes longer. When it moves once around the compact $u_1$
direction, it has undergone a monodromy, now wrapping the cycle indicated in purple.
\end{figure}

Consider a D4-brane wrapped along the $u_2$ direction, and sitting at a point in the $u_1$ direction. (As we
will discuss further below, we must also include a compensating source of $\overline{\rm D4}$-brane charge.)
This wrapped D4-brane contributes a potential energy in four dimensions of order
$\frac{L_{4B}}{(2\pi)^4(\alpha')^2 g_s}$, where $L_{4B}\sqrt{\alpha'}$ is the length of the cycle wrapped by the
D4-brane, $\alpha'$ is the inverse string tension and $g_s$ the ten-dimensional string coupling. At $u_1=0$,
this length is $L_{u_2}\sqrt{\alpha'}$, and as a function of $u_1$ it is given by
$L_{4B}=\sqrt{L_{u_2}^2+M^2L_x^2u_1^2}$. That is, as the D4-brane moves in the $u_1$ direction away from
$u_1=0$, it becomes heavier.  We will shortly derive the form of this curvature-induced potential as a function
of the canonically normalized field $\phi$ related to $u_1$.

We can now immediately see that the range of the field is not {\it geometrically} limited in the way it was for
a single D3-brane in a Calabi-Yau manifold \cite{FieldRange}. In the case considered in \cite{FieldRange}, the
approximate moduli space of the brane and the four-dimensional Planck mass are both determined by the volume of
the compactification.  In the present case, the four-dimensional Planck mass $M_{\rm P}$ is fixed as usual by
the volume of the compactification (in the absence of large species enhancements to it \cite{Nflation}, which we
will exclude in assessing the corrections below).  But the brane's field space is not constrained by the
compactification volume, as a result of the monodromies just described. Consider the motion of the D4-brane in
the $u_1$ direction. As the $u_1$ position of the D4-brane reaches $1$, the D4-brane has moved once around the
compact manifold. However, it now lies on the cycle $(y_1,y_2)=(M\lambda,\lambda)$ in the two-torus (note that
it still carries the same charge since the 1-cycle it wraps lies in the same homology class).  The approximate
moduli space of the brane lies on a subspace of the covering space of ${\cal N}_3$ -- the space obtained by
undoing the projection by $t_{u_1}$ in (\ref{translations}). Hence, at fixed four-dimensional Planck mass
$M_{\rm P}$, the geometrical moduli space of a single wrapped D4-brane is unlimited in field range.

We will seek a concrete model realizing this mechanism, on a product of two Nil 3-manifolds with a D4-brane
wrapped on a linear combination of their $u_2$ directions, whose motion in a linear combination of their $u_1$
directions describes inflation. (We will also consider a variant with motion along a linear combination of $u_1$
and $u_2$ directions.)  Starting out away from the minimum of the candidate inflaton potential corresponds to
beginning with a D4-brane wound up in the manner just described; as it unwinds by circling the $u_1$ directions
multiple times it approaches the minimum of its potential and eventually exits from inflation.

So far there are some similarities between our mechanism for a geometrically large field range and previously
proposed mechanisms such as \cite{Nflation,BLS}.  The fact that after moving once around the base circle in the
$u_1$ direction our brane lies on the $(M,1)$-cycle of the 2-torus described above means that there is some
sense in which it constitutes multiple or wrapped branes.  As noted in the previous works, a large field range
is not in itself sufficient to produce a consistent model of large-field inflation. We must check that the
potential energy carried by the inflaton not back react significantly on the geometry and destabilize the
compactification, and that all $\alpha'$ and quantum corrections to the inflaton potential are small or can be
tuned to preserve the smallness of the slow roll parameters everywhere on the inflaton trajectory.

In the previous examples, dynamical criteria of this sort happened to become important parametrically right at a
field range $\Delta\phi$ of order $M_{\rm P}$. In our case, we will find that enforcing these dynamical criteria
for control remains consistent with $\Delta\phi\gg M_{\rm P}$ in our Nil manifold compactification.  We will be
led to a window with fixed, small couplings and curvatures where our candidate model appears viable from the
theoretical and observational points of view.

Clearly this monodromy effect -- the fact that the D-brane approximate moduli space lives on a cover of the
compactification -- arises much more generally \cite{Albionmonodromy}\ than in the setup based on Nil
3-manifolds which we consider here.  In particular, there are similar monodromies for Sol 3-manifolds, where the
$SL(2,Z)$ transformation obtained in moving around the base circle is more general than $\tau\to\tau+M$.
However, as we will see, in that case the dynamical consistency criteria -- specifically the condition that the
inflaton potential not exceed the moduli-stabilizing curvature potential energy -- cleanly excludes this
possibility.\footnote{However, there exist stabilization mechanisms with larger potential energy -- such as
those coming from hyperbolic compactifications and/or supercritical ones -- as reviewed in \cite{Fluxreviews},
and these may accommodate a wider class of inflationary models.} It would be interesting to perform a systematic
study of models making use of this mechanism in more general compactifications such as
\cite{Albionmonodromy,Monodrofold}. We turn next to a systematic analysis of the D4-brane dynamics in our Nil
manifold case.

\subsection{Dynamics of wrapped D4-branes}

The effective action for the collective coordinates $X$ of our D4-brane in a given background, with dilaton
$\Phi(X)$, metric $G_{MN}(X)$, Neveu-Schwarz potential $B_{MN}(X)$, and RR potentials $C^{(p)}(X)$ is
\be\label{DfourSgen} {\cal S}_{\rm D4}=-\int \frac{d^5\xi}{(2\pi)^4(\alpha')^{5/2}}\,e^{-\Phi}
\sqrt{\det{(G_{MN}+B_{MN})\, \del_\alpha X^M\del_\beta X^N}} + {\cal S}_{CS} ~+~  loops \ee
This DBI action is self-consistently valid near solutions in which higher derivatives of the fields are
negligible, and we will estimate the size of corrections to it in later sections.  It is crucial in analyzing
(\ref{DfourSgen}) to plug in the correct solutions for $G_{MN},\Phi,$ and $B_{MN}$, including all effects of the
compactification \cite{KKLMMT,BDKMS,BDKM,Sequestering}\ and all relevant $\alpha'$ and loop effects.  In
particular, the compactification affects the Green's functions determining fields emanating from sources inside
it \cite{KKLMMT,BDKMS,BDKM,Sequestering}\ in a way that generically produces order 1 contributions to the
slow-roll parameters in brane inflation models.

The Chern-Simons action is given by
\be \label{DfourSCS} {\cal S}_{CS}= {1\over{(2\pi)^4(\alpha')^{5/2}}} \int
\left(\sum_p C^{(p)}\right)e^{-B}
e^{2\pi\alpha' F} \ee
where $F$ is the worldvolume gauge flux (which will play no role in our considerations).
${\cal S}_{CS}$ simplifies greatly in our situation, as follows.  The background form fields in \cite{ESdS}\
preserve the Lorentz symmetry in four dimensions.  Therefore (\ref{DfourSCS})
can  only get contributions from $C^{(5)}_{0123{u_2}}$, with legs along the five worldvolume dimensions of the
D4-brane. In the original moduli-stabilized compactification, there is no such background field  -- the
solutions in \cite{ESdS}\ did not make use of an $F_6$ flux in these directions (or its dual internal $F_4$
flux).  However as mentioned above, we must have a source of $\overline{\rm D4}$-brane charge in the
compactification in order to satisfy Gauss' law for $C^{(5)}$.  We will include this source in the form of flux
below in \S\ref{sec:corrections}, arguing that its homogeneity suppresses its contributions to the slow-roll
parameters in our model.


%
%

The background configurations of $\Phi,G_{MN}$, and $B_{MN}$ appearing in (\ref{DfourSgen}) are given by the nil
geometry (\ref{nilgeom}) plus contributions from the other ingredients required to stabilize the moduli (which
in the case of the construction \cite{ESdS}\ consist of fivebranes, O6-planes, zero-form and six-form fluxes
$m_0$ and $F_6$, and discrete Wilson lines built from the $B$ field) plus more general $\alpha'$ and loop
corrections. We will start by analyzing the contributions to the D4-brane dynamics coming from the Nil geometry
itself, deriving a curvature-induced potential $V_{\cal R}(\phi)$ for the canonically normalized field $\phi$
corresponding to the D4-brane's motion in the $u_1$ direction.  A crucial requirement we must impose is that
this candidate inflaton potential $V_{\cal R}(\phi)$ not destabilize the moduli or back react strongly on the
geometry (or vice versa), and we find that this condition is straightforward to satisfy for a large range of the
field $\phi$. In the relevant range, we find that the potential $V_{\cal R}(\phi)$ is well approximated by a
power law potential $V(\phi)\propto \phi^{2/3}$, leading to specific predictions for two basic CMBR observables,
$r$ (the tensor to scalar ratio) and $n_s$ (the tilt of the spectrum) which lie in an observationally accessible
regime, near the central value for these quantities according to existing data.

We will then address in \S\ref{sec:corrections}\ the other contributions from the remaining ingredients in the
moduli stabilization mechanism \cite{ESdS}, and from $\alpha'$ and loop corrections to the background, arguing
that for appropriate configurations of ingredients, these need not destabilize the resulting inflaton
trajectory.

\subsection{The curvature-induced inflaton potential}

We consider type IIA string theory compactified to 4d on an orientifold of a product of two identical Nil
manifolds ${\cal N}_3\times \tilde {\cal N}_3$ ~\cite{ESdS}. The Nil 3-manifold ${\cal N}_3$ was described
above, with metric
\be\label{Nilgeom} \frac{ds^2}{\alpha'}=L_{u_1}^2du_1^2+L_{u_2}^2du_2^2+L_x^2(dx'+M u_1 du_2)^2 \ee
and we have an isomorphic metric for the second Nil 3-manifold factor, with corresponding coordinates $\tilde
u_1,\tilde u_2,\tilde x'$.

Assume now a D4-brane wrapped along the $u_2$-direction (or a combination of the $u_2$ and $\tilde u_2$
directions) and watch it as it moves along $u_1$ (or similarly a combination of the $u_1$ and $\tilde u_1$
directions). As explained above, the energy density of the D4-brane grows monotonically with increasing $u_1$
even as the brane moves along $u_1$ multiple times through the fundamental domain of the geometry.  In order to
analyze the resulting candidate inflaton dynamics, it is most convenient to transform to the canonically
normalized scalar field $\phi$ corresponding to the motion of the D4-brane in the $u_1$ direction.

We start with the DBI action (\ref{DfourSgen}) for a single D4-brane, with zero $B_{(2)}$ in the worldvolume
directions, and also vanishing worldvolume gauge field strength $F_{(2)}$.
%
%
The four-dimensional Planck mass $M_{\rm P}$ is related to the string tension $1/\alpha'$ via
\be\label{alphapr} \frac{1}{\alpha'}={(2\pi)^7\over 2}g^2M_{\rm P}^2\quad{\rm with:}\quad
g^2=\frac{g_s^2}{L^6/2} \ee
the 4-dimensional string coupling defined from the reducing the type IIA supergravity action $S_{\rm
IIA}=-1/2\kappa^2\int d^{10}x\sqrt{-g}e^{-2\phi}R+\ldots$ on the orientifold of ${\cal N}_3\times \tilde{\cal
N}_3$ with volume $\CV=L^6/2$.

\begin{figure}[t]
\begin{center}
\includegraphics[width=16cm]{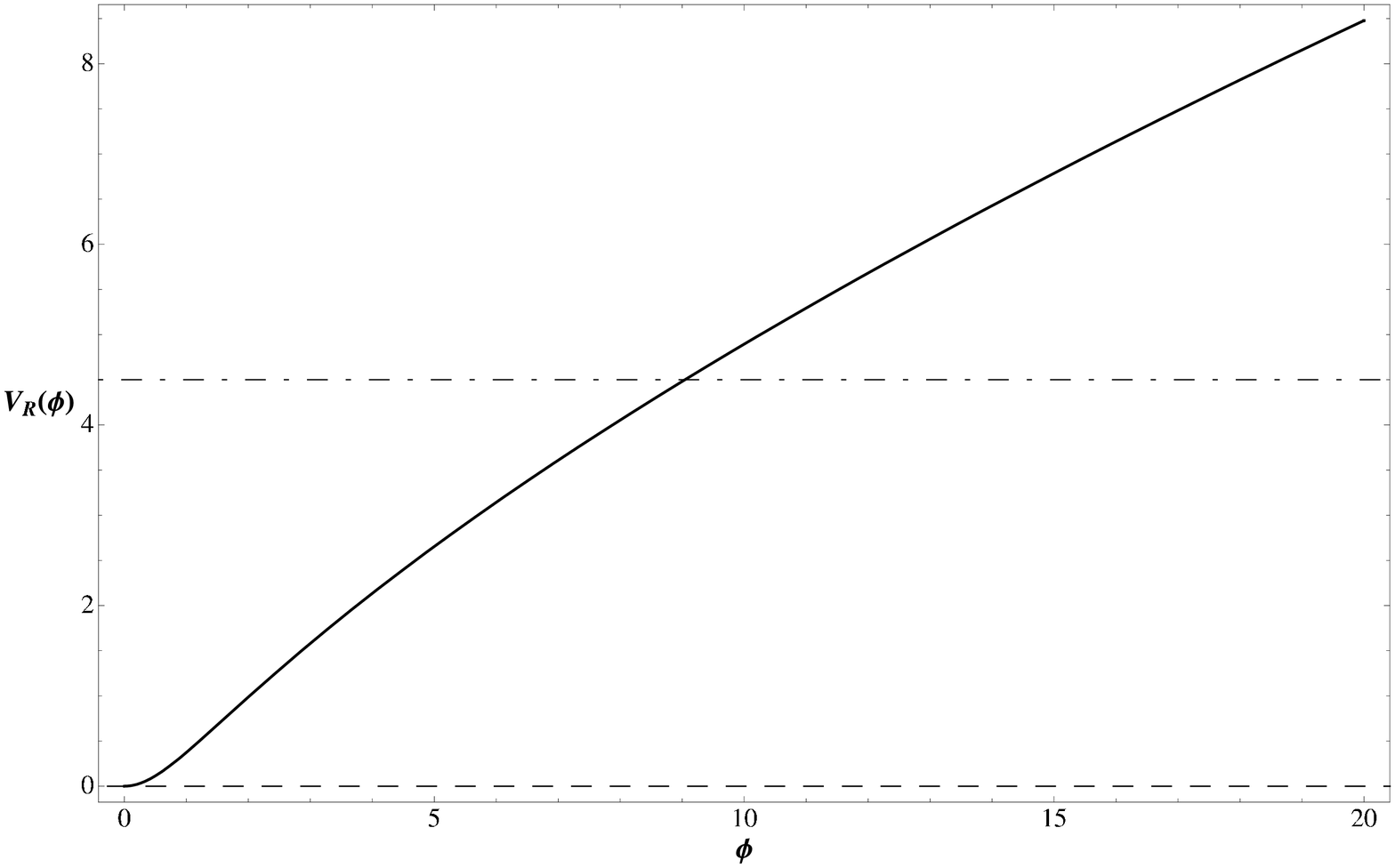}
\end{center}
\refstepcounter{figure}\label{Vplot1}
\vspace*{-.2cm} {\bf Figure~\ref{Vplot1}:} Solid black: The potential $V_{\cal R}(\phi)$ in arbitrary units for
the canonically normalized field $\phi$ (in Planck units) corresponding to the D4-brane position $u_1$.  The
horizontal (dash-dot) line indicates the scale of the moduli potential, and in the case displayed here the model
is minimally tuned to avoid destabilizing the moduli at a distance in field space at about $\phi\simeq 9M_{\rm
P}$ which is the minimal field range necessary to get 60 efolds of slow-roll inflation.  This corresponds to a
minimally tuned anisotropy parameter $\beta\sim 0.04$ (see below).
\end{figure}

The `radial' modulus $L$ is defined in terms of the size moduli for the $u_i,\tilde u_i,x,\tilde x$ directions
as $L^3=L_{u_1}L_{u_2}L_x=L_{\tilde u_1}L_{\tilde u_2}L_{\tilde x}$. As $u_1$ and $u_2$ are interchangeable in
their respective roles before the introduction of the D4-brane, we define a common scale $L_u$ such that
$L^3=L_u^2L_x$ and parametrize the anisotropy of the Nil manifold through the modulus
\be\label{betadef} \beta\equiv \frac{L_{u_2}}{L_{u_1}}=\frac{L_u^2}{L_{u_1}^2}=\frac{L_{u_2}^2}{L_u^2} \ee
which can be tuned by an appropriate choice of flux and brane quanta~\cite{ESdS}, as we will discuss further
below. (Ultimately a 1 percent fine-tune will be needed.)  In terms of these quantities, we obtain the following
action for the fluctuations of the brane position $u_1$
\bea\label{D4a}
S_{\rm D4}[u_1]&=&-\frac{1}{(2\pi)^4 g_s \alpha'^{5/2}}\int_{{\cal M}_4\times u_2}\d^5\xi\sqrt{-g_4g_{u_2u_2}\left(1-\alpha'g_{u_1u_1}\dot u_1^2\right)}\nn\\
&=&-\frac{1}{(2\pi)^4 g_s \alpha'^{2}}\int_0^1 du_2\int_{{\cal M}_4}\d^4\xi\sqrt{-g_4}\sqrt{(\beta L_u^2+L_x^2M^2 u_1^2)\left(1-\alpha'\frac{L_u^2}{\beta}\dot u_1^2\right)}\nn\\
&=&-\frac{1}{(2\pi)^4 g_s \alpha'^2}\int d^4x\sqrt{-g_4}\sqrt{(\beta L_u^2+L_x^2M^2
u_1^2)\left(1-\alpha'\frac{L_u^2}{\beta}\dot u_1^2\right)}\quad. \eea
Upon expanding this action up to two derivatives we arrive at
\beq\label{D4b} S_{\rm D4}=\frac{1}{(2\pi)^4 g_s \alpha'^2}\int d^4x\sqrt{-g_4}\left(\beta^{-1}L_u^2\sqrt{\beta
L_u^2+L_x^2M^2 u_1^2}\alpha'\frac{\dot u_1^2}{2}-\sqrt{\beta L_u^2+L_x^2M^2 u_1^2}+\ldots\right)\;. \eeq
This action describes the dynamics of a non-canonically normalized scalar field with a positive definite
potential. To exhibit the dynamics of $u_1$ we pass to a canonically normalized scalar field $\phi$ with kinetic
term $S=\int d^4x\sqrt{-g_4}\dot\phi^2/2$. In terms of the D4-brane position $u_1$ we then have
\beq \label{cannormreln} \dot\phi=\phi'(u_1) \dot u_1\quad \Leftrightarrow\quad
\frac{\dot\phi^2}{2}=\phi'^2\frac{\dot u_1^2}{2} \eeq
and thus
\bea\label{cannorm} \phi'(u_1)&=&\frac{L_u^{3/2}\beta^{-1/4}}{(2\pi)^2\sqrt{g_s\alpha'}}
\left(1+\frac{M^2L_x^2}{\beta L_u^2}\,u_1^2\right)^{1/4}\nn\\
\Rightarrow\quad\phi&=&\frac{L_u^{3/2}\beta^{-1/4}}{3(2\pi)^2\sqrt{g_s\alpha'}}u_1
\left[F_{1,\frac{1}{2},\frac{3}{4},\frac{3}{2}}^2\left(-\frac{M^2L_x^2}{\beta
L_u^2}\,u_1^2\right)+2\left(1+\frac{M^2L_x^2}{\beta L_u^2}\,u_1^2\right)^{1/4}\right] \eea
where $F_{p,q,r,s}^2(x)$ denotes a hypergeometric function. In terms of this canonically normalized scalar field
$\phi$ the D4-brane action to quadratic order in the derivatives then reads \beq\label{D4c} S_{\rm D4}=\int d^4
x\sqrt{-g_4}\left[\frac{1}{2}\,\dot\phi^2-V_{\cal R}(\phi)\right] \eeq with \beq\label{pot1} V_{\cal
R}(\phi)=\frac{\sqrt\beta L_u}{(2\pi)^4g_s\alpha'^2}\sqrt{1+\frac{M^2L_x^2}{\beta L_u^2} \, u_1^2(\phi)} \eeq
where $u_1(\phi)$ is to be read as the inverse function of eq. (\ref{cannorm}). This inversion we could not do
analytically.  However, it is possible to invert $\phi(u_1)$ numerically which results in $V_{\cal R}(\phi)$ as
displayed in Fig.~\ref{Vplot1}.

%
One can clearly see in Fig.~\ref{Vplot1} that there is a regime $V_{\cal R}(\phi)\propto \phi^2$ close to the
origin and that at large super-Planckian field values the potential grows slower than$~\phi$.  One can determine
the behaviour of $V_{\cal R}(\phi)$ in these two regimes analytically by appropriately expanding
eq.~(\ref{cannorm}) or equivalently the square root in eq.~(\ref{D4b}) and then determining the canonically
normalized field $\phi$ from that expansion.

We find that for \beq
 u_1< u_{1,\rm crit}\sim \frac{1}{M}\sqrt\beta\,\left(\frac{L}{L_x}\right)^{3/2}
\eeq the action eq.~(\ref{D4b}) expands as \beq\label{D4d} S_{\rm D4}=\frac{\sqrt\beta L_u}{(2\pi)^4 g_s
\alpha'^2}\int d^4x\sqrt{-g_4}\left(\frac{L_u^2}{\beta}\,\alpha'\frac{\dot u_1^2}{2}-\frac{L_x^2M^2}{2\beta
L_u^2} u_1^2\right)\;. \eeq This becomes canonically normalized for \beq\label{phican1}
\phi=\frac{L_u^{3/2}}{(2\pi)^2\sqrt{g_s\alpha'}\beta^{1/4}}\,u_1 \eeq yielding \beq\label{D4dcan} S_{\rm
D4}=\int d^4 x\sqrt{-g_4}\left(\frac{1}{2}\,\dot\phi^2-\frac{m^2}{2}\,\phi^2\right) \eeq with
\beq\label{phimass} m^2=\frac{M^2}{\alpha'}\frac{L_x^4}{L^6}\quad. \eeq The boundary value $u_{1,\rm crit}$
where this expansion breaks down corresponds to a value of $\phi$
\beq\label{phicrit} \frac{\phi_{\rm crit}}{M_{\rm P}}\sim (2\pi)^{3/2}\beta^{1/4}{\sqrt{g_s}L^{3/4}\over{M
L_x^{9/4}\sqrt{2}}} \eeq
Here we have plugged in the definition of $\alpha'$ from eq.~(\ref{alphapr}) to express $\phi_{\rm crit}$ in 4d
Planck units.  In order to determine from (\ref{phicrit}) whether (and under what conditions) the $m^2\phi^2$
behavior persists into the regime of $\phi_{\rm crit}> M_{\rm P}$, one needs a concrete stabilization mechanism
for $g_s, L,L_x,$ and $\beta$.  We will see in the construction \cite{ESdS}\ that $\phi_{\rm crit}$ is
sub-Planckian parametrically as we take appropriate flux quantum numbers large, while maintaining the
stabilization of the moduli.

In fact, we can derive this conclusion -- that the $m^2\phi^2$ regime of the potential (\ref{pot1}) never
applies self-consistently for super-Planckian field range -- in a more general way. It follows from the
requirement that our inflaton potential energy not exceed the scale of the potential energy ${\cal U}_{mod,{\cal
R}}$ used to stabilize the moduli.  In any model on Nil manifolds, the contribution to the moduli potential
coming from the negative scalar curvature ${\cal R}$ of the compactification is of the order
\be\label{Umodcurv} {\cal U}_{mod,\cal R}\sim M_{\rm P}^4 {(2\pi)^7\over 4}g^2 {M^2 L_x^4\over L^{6}}  \ee
Meanwhile, the potential $m^2\phi^2$ from (\ref{D4dcan})(\ref{phimass}) is
\be\label{VmassPlanck} V_{\cal R}(\phi<\phi_{\rm crit})={1\over 2}m^2\phi^2\sim \left({\phi\over M_{\rm
P}}\right)^2M_{\rm P}^4{(2\pi)^7\over 4}g^2 {M^2 L_x^4\over L^{6}}\sim \left({\phi\over M_{\rm P}}\right)^2
{\cal U}_{mod,\cal R}  \ee
From this we see that the requirement of consistency with moduli stabilization prevents a super-Planckian field
range for $\phi$ within the $m^2\phi^2$ regime of the potential (\ref{pot1}).  This rules out single-field
slow-roll inflation in the $m^2\phi^2$ regime of our potential -- the slow roll parameters $\epsilon={M_{\rm
P}^2\over 2}({V'\over V})^2$ and $\eta=M_{\rm P}^2 {V''\over V}$ are at least of order 1 for $\phi\le M_{\rm
P}$.  Of course this does not rule out the possibility in general --for example, with a higher-scale moduli
potential, arising in the less supersymmetric constructions among those reviewed in \cite{Fluxreviews},
it may be possible to accommodate $m^2\phi^2$ inflation in a different setup.

\subsubsection{Sol manifolds}
\label{subsubsec:sol}

Before continuing with our main example, let us briefly mention a similar result for Sol manifolds.  Sol
3-manifolds constitute a more general class of twisted tori where the $SL(2,Z)$ transformation made in going
around the base circle is more generic than a $\tau\to\tau+M$ transformation.  These arise by compactifying the
{\it sol geometry}
\be\label{solgeom} ds^2=L_1^2dx_1^2e^{2z}+L_2^2dx_2^2 e^{-2z}+L_z^2 dz^2 \ee
In this case, the curvature is $-2/L_z^2$ and the volume of each 3-manifold factor is $L^3=L_1L_2L_z$, leading
to a contribution to the moduli potential of order
\be\label{solmodcurv} {\cal U}_{{\cal R},sol}\sim M_{\rm P}^4{{(2\pi)^7g_s^2}\over{4 L_z^2L^6}} \ee
For a D4-brane wrapped along some generic linear combination of the $x_1$ and $x_2$ directions, and moving in
$z$, we obtain from the DBI action with metric (\ref{solgeom}) an effective action which both for large and
small $z$ has a potential of order
\be \label{solpot} V_{{\cal R},sol}\sim m^2\phi^2\sim {1\over {L_z^2\alpha'}}\phi^2 \sim {\cal U}_{{\cal
R},sol}\left({\phi\over M_{\rm P}}\right)^2\ee
in terms of the canonically normalized field $\phi$, where in the last step we used (\ref{alphapr}).

Thus in both Nil and Sol manifolds, an $m^2\phi^2$ potential is not consistent with moduli stabilization,
assuming that the curvature-induced potential is one of the leading terms in ${\cal U}_{mod}$.  Again, it is
possible that string compactifications with higher-scale potentials -- such as those arising on more generic
hyperbolic compactifications and/or from $D>10$ string theory -- could accommodate an $m^2\phi^2$ inflationary
potential, with the moduli stabilizing potential arising at a higher scale. However in absence of that, we
obtain a clean parametric constraint on $m^2\phi^2$ inflation in twisted tori for which the scalar curvature
contributes to moduli stabilization as in \cite{ESdS}.

\subsubsection{Our main example:  large $u_1$ and the $\phi^{2/3}$ potential}

Because we cleanly ruled out the $m^2\phi^2$ regime, let us focus on the opposite regime where $\phi\gg\phi_{\rm
crit}$. In this case, the action eq.~(\ref{D4b}) expands as \beq\label{D4e} S_{\rm D4}=\frac{1}{(2\pi)^4 g_s
\alpha'^2}\int d^4x\sqrt{-g_4}\left(\frac{1}{2}\,\frac{M L_x L_u^2}{\beta}\,\alpha'\,u_1\,\dot u_1^2-L_x M
u_1\right)\;. \eeq This becomes canonically normalized for \beq\label{phican2} \phi=\frac{M^{1/2} L_u
L_x^{1/2}}{6\pi^2\sqrt{g_s\alpha'}\beta^{1/2}}\,u_1^{3/2} \eeq yielding \beq\label{D4ecan} S_{\rm D4}=\int d^4
x\sqrt{-g_4}\left(\frac{1}{2}\,\dot\phi^2-\mu^{10/3}\,\phi^{2/3}\right) \eeq with
\beq\label{lambda}
\mu^{10/3}=\left(\frac{3}{2}\right)^{2/3}(2\pi)^{-8/3}\,\frac{M^{2/3}\beta^{1/3}}{\alpha'^{5/3}g_s^{2/3}}\,
\frac{L_x}{L}\quad. \eeq
Altogether, we see that for $\phi\gg\phi_{\rm crit}$ we have a potential for $\phi$ (coming from the curvature
of the compactification which affects the D4-brane action) given by a fractional power \be\label{curvpot}V_{\cal
R}(\phi)\propto \phi^p,\,p=2/3 \ee Single-field slow roll inflation based on any power law potential with power
$p$ of order 1 requires a super-Planckian field range in order to suppress the slow-roll parameters $\epsilon =
{M_{\rm P}^2\over 2}({V'\over V})^2$ and $\eta=M_{\rm P}^2 {V''\over V}$.  So far, our system thus provides a
candidate for large field chaotic inflation with a fractional power-law potential.\footnote{The term ``chaotic
inflation" follows the title of the work \cite{Linde:1983gd}\ which introduced the first example. But see
\cite{Vilenkinchaos}\ for an analysis of the ambiguity in terminology which has resulted in the literature,
since this term may refer only to the assumption of chaotic initial conditions.  We cannot implement the
suggestion of \cite{Vilenkinchaos}\ and call our model ``inflation with an unbounded potential" simply because
the moduli-stabilizing potential will always cut it off at some point in field space.}

There are many self-consistency conditions and observational constraints which must now be imposed in order to
assess this possibility.  This will occupy the bulk of the remainder of the paper.  We will find a reasonably
natural viable regime where (\ref{curvpot}) seems to apply to good approximation, noting a few subtleties along
the way. We will then review the observational predictions of single-field slow roll inflation governed by our
power-law potential (\ref{curvpot}).  Before turning to that, let us pause to note another version of this
mechanism, which gives a different power-law potential in a similar way.

\subsubsection{A potential variant with $\tilde V(\tilde\phi)\propto \tilde\phi^{2/5}$ }
\label{subsubsec:variant}

In our product of two Nil 3-manifolds, we could consider a D4-brane wrapped on say the $u_2-\tilde u_2$
direction, while moving in a linear combination $u_B$ of $u_1,\tilde u_1$, and $u_2+\tilde u_2$ directions.  In
this case, the term $L_x^2M^2(u_1^2du_2^2+\tilde u_1^2d\tilde u_2^2)$ in the metric would lead to a contribution
to the kinetic energy for $u_B$ of the form
\be\label{tildemodel}\tilde {\cal L}_{kin}\sim \dots +{1\over{(2\pi)^4g_s\alpha^{'2}}}\sqrt{\beta
L_u^2+L_x^2M^2u_B^2}(L_x^2M^2u_B^2\dot u_B^2). \ee
In this case, for large $u_B$ the canonically normalized field $\tilde\phi$ would satisfy (using the analogue of
(\ref{cannormreln})) $u_B\propto\tilde\phi^{2/5}$, leading to a potential $\tilde
V(\tilde\phi)\propto\tilde\phi^{2/5}$.  We will continue to analyze the case with the $\phi^{2/3}$ potential in
detail; the $\tilde\phi^{2/5}$ case can be analyzed similarly.


\subsection{Consistency with Moduli Stabilization and Background Geometry}
\label{subsec:selfconsistency}

In this subsection, we will analyze several basic self-consistency criteria.  First, we will impose the
condition that the inflaton potential energy not destabilize the moduli.  Next, we will show that the small
$\phi$-dependent shifts in moduli which do arise do not destabilize the inflaton trajectory.  Finally, we will
analyze and bound the back reaction of our D4-brane in the ten-dimensional description.

\subsubsection{Condition $V_{\cal R}(\phi)<{\cal U}_{mod,{\cal R}}$}

A crucial first condition for the above mechanism of large-field inflation to succeed in concrete model is that
the process of inflation at $\phi>M_{\rm P}$ not destabilize the moduli (a condition which immediately rules out
the $m^2\phi^2$ regime of our potential, as discussed above). The moduli potential or, more precisely, the
height of the barriers protecting the moduli $L,L_x,L_u$ from run-away scales as indicated in
(\ref{Umodcurv})~\cite{ESdS}.\footnote{We suppress dependence on the angular moduli discussed in \cite{ESdS}.
These angles have positive mass squareds in the rectangular configuration we are considering, at fixed $L,L_x,$
and $\beta$. However, we have not explicitly computed the off-diagonal elements in the mass matrix between the
angular moduli and the others.  If these are large enough, tachyonic directions could develop; this depends on
order 1 coefficients in the potential and its derivatives. A cursory look at the terms which contribute to the
mass matrix suggests that these off-diagonal elements are likely to be tuneable to be numerically smaller than
the diagonal contributions. In particular, there are many ingredients one can add which are of the same order or
less than the original moduli potential, and which push on the angles in different directions.  However this --
and the similar problem of determining the off-diagonal elements in the mass matrix mixing the lightest KK modes
with the moduli in the simplest version of the setup described in \cite{ESdS}\ -- remains to be worked out
explicitly.} Introducing a source of potential energy which dominates over those used in \cite{ESdS}\ would
generically remove the local de Sitter minimum found there. Thus, the most basic condition for preserving moduli
stabilization is
\beq\label{stabcond} V_{\cal R}(\phi)<{\cal U}_{mod,{\cal R}}\quad{\rm for:}\quad \phi<\phi_{\rm max}\quad. \eeq
Applying this condition now to the $\phi^{2/3}$ regime applicable for $\phi\gg\phi_{\rm crit}$, we can start by
determining whether observationally viable large field chaotic inflation can coexist with moduli stabilization
at this basic level.

We find
\beq\label{phimax2} \phi_{\rm max}\sim M_{\rm P} (2\pi)^{21/2}\frac{M_{\rm P}^5}{\mu^5}\,\frac{M^3g^3
L_x^6}{8L^9}\quad. \eeq
We note that as $\mu^{10/3}\propto\beta^{1/3}$ this implies that in this regime we have
\beq \phi_{\rm max}\propto \beta^{-1/2}\quad, \eeq
suggesting that $\beta\ll 1$ may provide a parametric regime where $\phi_{\rm max}$ grows large -- this is
intuitive since suppressing $\beta$ corresponds to shrinking the cycle wrapped by our D4-brane, while also
expanding the direction $u_1+\tilde u_1$ in which it moves.  However, we will find compensating constraints on
$\beta$ which prevent us from tuning it to be arbitrarily small.

The value of $\phi_{\rm max}$ depends on the stabilized values of the moduli. It is now worth plugging in the
results for the moduli in terms of flux quantum numbers in the construction \cite{ESdS}, type IIA on a product
${\cal N}={\cal N}_3\times{\cal \tilde N}_3$ of Nil three-manifolds with six-form flux $F_6$ satisfying the
quantization condition $\int_{\cal N} F_6=K$, with
\be\label{sixform} K={1\over\sqrt{2}}(2\pi)^5 f_6 \sim 7\times 10^3 f_6 \ee
in terms of the integer flux quantum number $f_6$.  In this construction, the VEVs of the moduli in their dS
minimum scale in terms of the basic topological and flux quantum numbers $M,K$ as
\beq L=c_L\cdot K^{1/6}\quad,\quad L_x=\frac{c_{L_x}}{M^{1/2}}\quad, \quad g={{\hat g}\over K} \eeq
$c_L,c_{L_x}$ here are numerically constant, which for a generic Nil manifold compactification will be of ${\cal
O}(1)$.

In terms of these quantities, we find for $\phi_{\rm crit}$ and $\phi_{\rm max}$
\bea \frac{\phi_{\rm crit}}{M_{\rm P}}&\sim&(2\pi)^{3/2}\beta^{1/4}{\hat g}^{1/2}
\left(\frac{M}{K}\right)^{1/8}\left(\frac{c_L}{c_{L_x}}\right)^{9/4}\nn\\
\frac{\phi_{\rm max}}{M_{\rm P}}&\sim&\frac{1}{3}\,\beta^{-1/2}\left(\frac{K}{M}\right)^{1/4}
\left(\frac{c_L}{c_{L_x}}\right)^{-9/2}{1\over {(2\pi)^3{\hat g}}}\\
\eea
These relations go in the right direction for our mechanism, as the ratio $M/K$ is much smaller than one in even
the simplest version of the construction \cite{ESdS}.  In some ways, this construction becomes better controlled
(in the sense that the KK mass scale is parametrically larger than the moduli mass scale) in the regime
$K/M\to\infty$, though arranging this requires extra ingredients which complicate the analysis. In this
regime,\footnote{and also for $\beta\ll 1$, which is limited by competing effects we will soon discuss}
$\phi_{\rm crit}$ decreases and $\phi_{\rm max}$ increases.  In all versions, we find ourselves squarely in the
regime $V_{\cal R}\propto\phi^{2/3}$. See Fig.~\ref{Vplot1} for a graphical depiction of the candidate inflaton
potential and the moduli potential, showing the finite but super-Planckian range of $\phi$ consistent with
moduli stabilization.

Since we are including the degree of freedom $\beta$, we must also keep track of its effects on the moduli
stabilization mechanism.  The ingredients which contribute leading terms to the moduli potential must be
arranged so as to produce the value of $\beta\sim L_2/L_1$ used for inflation.  We will be led to consider, for
example, KK fivebranes on the cycle generated by the translations $t_{u_1} t_{{\tilde u}_1} t_{u_2}^{1/b}
t_{{\tilde u}_2}^{1/b}$, with $b$ taken to be of order $\beta$ so as to stabilize $L_{u_2}/L_{u_1}$ at the value
$\beta$. This introduces a factor of $1/\sqrt{\beta}$ in the corresponding KK5-brane contribution to the
potential energy relative to the expression in eq.~(3.22) of \cite{ESdS}\ (which pertained to the isotropic
arrangement $L_{u_1}=L_{u_2}$).  Now we have
\be\label{UKK} {\cal U}_{KK5}\propto {1\over\sqrt{\beta}}M_{\rm P}^4 g^2 {M L_x^{5/2}\over{L^{9/2}}}  \ee
Because of this, $\beta$ cannot be reduced arbitrarily, since the terms in ${\cal U}_{mod}$ were already all of
the same order in the construction \cite{ESdS}. In a sample example to be discussed below, we will find that we
need only a modest tune of $\beta\sim 0.04$.  However, eternal inflation would require a much stronger tune of
$\beta$, which would need to be reevaluated vis a vis moduli stabilization.

Let us finally note here that the condition eq.~(\ref{stabcond}) necessary for avoiding moduli destabilization
implies the same relationship between the inflationary Hubble parameter and the gravitino mass
\beq\label{gravitinoscale} H_{\rm inf}\le m_{3/2} \eeq
as discussed in the context of the KKLT-style IIB de Sitter vacua in~\cite{KLgravitino}. In our present case,
this follows from the fact that the scale of supersymmetry breaking in a typical Nil manifold compactification
is at or above the curvature scale:  $m_{3/2}\ge \sqrt{{\cal R}}$.  So since ${\cal U}_{mod,{\cal R}}\sim M_{\rm
P}^2(-{\cal R})$, (\ref{stabcond}) implies
\be\label{curvalpha} V_{\cal R}\sim M_{\rm P}^2 H_{\rm inf}^2\le M_{\rm P}^2m_{3/2}^2  \ee
%

%
%

\subsubsection{Moduli Shifts and the Inflaton potential}

The moduli are not destabilized by our additional contribution $V_{\cal R}$ to the potential energy, but they do
shift slightly due to its presence.  Let us analyze the effect of this on our candidate inflaton potential.
Schematically the potential is of the form
\be\label{potLphi} {\cal U}_{\rm tot}\sim {\cal U}_{mod}(L e^{{\sigma}\over{M_{\rm P}}},\dots)+V_{\cal
R}(\phi,Le^{{\sigma}\over{M_{\rm P}}},\dots) \ee
where we keep track of the dependence on both the inflaton $\phi$ and on the moduli (represented by $L
e^{{\sigma}\over{M_{\rm P}}}$,\dots).  Here $L$ denotes the stabilized value of the modulus in the absence of
the inflaton potential, and $\sigma$ is the canonically normalized scalar field describing the deformation of
the modulus away from this value.

In particular, $\del_\sigma |_{\sigma=0} {\cal U}=0$. Taylor expanding the full potential about $\sigma=0$, we
obtain (suppressing the dependence on the other moduli $\dots$, which work similarly)
\be\label{Taylorpot} {\cal U}_{\rm tot}\sim {\cal U}_{mod}(L)+{1\over 2}\del_\sigma^2{\cal
U}_{mod}(L){\sigma^2\over M_{\rm P}^2} + V_{\cal R}(\phi, L)+\del_\sigma V_{\cal R}(\phi, L){\sigma\over M_{\rm
P}}+ {1\over 2}\del_\sigma^2 V_{\cal R}(\phi,L){\sigma^2\over M_{\rm P}^2} \ee

Now $V_{\cal R}$ and each individual term in the moduli potential is proportional to a power of $Le^{\sigma\over
M_{\rm P}}$.  So the derivatives of each term with respect to $\sigma$ scale the same parametrically -- they
depends on the same powers of $L,\dots$ -- as the term itself. Note that although the de Sitter minimum of the
moduli potential is tuned to lie near zero vacuum energy:  ${\cal U}_{mod}(L)\ll {\cal U}_{mod,{\cal R}}(L)$,
the second derivative $\del_\sigma^2{\cal U}_{mod}(L)$ is not tuned to be small.  It is of the same order as a
typical term in the moduli potential, $\sim {\cal U}_{mod,{\cal R}}$.

Putting this together, we have from (\ref{Taylorpot}) a small shift in the moduli
\be\label{modshift} {\sigma\over M_{\rm P}}\sim {{\del_\sigma V_{\cal R}(\phi,L)}\over{\del_\sigma^2{\cal
U}_{mod}(L)+\del_\sigma^2 V_{\cal R}(\phi,L)}}\sim {{V_{\cal R}(\phi,L)}\over{{\cal U}_{mod,{\cal R}}(L)}} \ee
Plugging this back into the potential (\ref{Taylorpot}) we have (for some order one constant $c_{L\phi}$)
\be\label{modcorrpot} {\cal U}_{tot}\sim {\cal U}_{mod}(L)+ V_{\cal R}(\phi, L)+ c_{L\phi}{V_{\cal
R}(\phi,L)^2\over{{\cal U}_{mod,{\cal R}}(L)}} \ee
This gives a small change in the functional form of the inflaton potential, shifting the slow-roll parameter
$\eta=M_{\rm P}^2 {{\del_\phi^2 V}\over V}$ by a term of order $\le \eta$ itself
\be\label{shifteta}\Delta\eta\sim \eta {V_{\cal R}(\phi,L)\over{{\cal U}_{mod,{\cal R}}}} \ee
Thus these small shifts of the moduli do not destabilize inflation, and only for the case $V_{\cal R}\sim
U_{mod,{\cal R}}$ saturating (\ref{stabcond}) could they contribute significantly to the tilt of the power
spectrum (which depends on $\eta$ in a way we review below).

\subsubsection{Back reaction of branes in $10d$}
\label{subsubsec:backreaction}

Deep in our regime where $V_{\cal R}(\phi)\propto\phi^{2/3}$, the D4-brane wrapped on the (M,1) cycle of the
torus traced out by $x',u_2$ constitutes to good approximation a set of $Mu_1$ branes wrapped in the $x$
direction, spaced evenly in the $u_2$ direction. As mentioned above in \S\ref{sec:dynamics}, in this sense we
are considering a multiply wrapped brane. As discussed in previous works such as \cite{BLS}, multiple and/or
wrapped branes -- which help extend the field range even for Calabi-Yau compactifications -- can lead to
significant back reaction, and we should check this in our case. In our discussion of loop corrections below, we
will also need the number of species introduced by our effectively multiply wrapped brane, in assessing the
strength of e.g. the renormalization of $M_{\rm P}$ \cite{Nflation}.

First, note that using the fact from (\ref{D4e}) that $V_{\cal R}\sim {L_xMu_1\over{(2\pi)^4g_s(\alpha')^2}}$,
and using (\ref{alphapr}) and (\ref{Umodcurv}), we can express the number of windings $Mu_1$ of our brane around
the $x$ direction as
\be\label{windingN} N_w \equiv Mu_1\sim {V_{\cal R}\over {\cal U}_{mod,{\cal R}}} {2 L_x^3M^2\over {(2\pi)^3
g_s}} \ee
We must check whether this multiply wound brane still constitutes a probe of the geometry, as we have assumed in
writing its DBI action (\ref{D4a}).

Let us dimensionally reduce on the $u_2,x,\tilde x$ directions, and determine the core size of our wrapped D4
brane in the remaining $\tilde u_2,u_1$, and $\tilde u_1$ directions.  We can estimate this as follows.  The
gravitational potential in the 3 directions $\tilde u_2,u_1,\tilde u_1$ is given by
\be\label{gravpot} \Phi_{grav}\sim {G_7 V_{\cal R}\over |\vec r|} \ee
where $G_7\sim (2\pi)^4{g_s^2(\alpha')^{5/2}\over L_x^2L_{u_2} }$ is the Newton constant in the remaining seven
dimensions after the dimensional reduction along $u_2,x,\tilde x$, and $|\vec r|$ is the proper distance from
the source D4-brane. Here we work in the flat space approximation (which will be valid if we arrange that the
core size is smaller than the curvature radii), using the BPS formula for the brane tensions which is a good
approximation for low curvature. From (\ref{gravpot}) we can read off the the core size of our brane:
\be\label{coresize} r_c\sim G_7V_{\cal R} \ee

If the core size $r_c$ is smaller than $L_{u_1}$ and the curvature radius $r_{\cal R}$, then the brane is a good
probe as far as motion in the $u_1$ direction goes (as we have been assuming).  First, note that
\be\label{coreLtwo} {r_c\over{L_{u_2}\sqrt{\alpha'}}}\sim {V_{\cal R}\over {\cal U}_{mod,{\cal R}}}{{ L_x^3
M^2}\over{\beta L^3 (2\pi)^3}}\sim {V_{\cal R}\over {\cal U}_{mod,{\cal R}}} {1\over{\beta (2\pi)^3}}
\sqrt{{M\over K}} \ee
Correspondingly,
\be\label{coreLone} {r_c\over{L_{u_1}\sqrt{\alpha'}}}\sim {V_{\cal R}\over {\cal U}_{mod,{\cal R}}}{{ L_x^3
M^2}\over{ L^3 (2\pi)^7}}\sim {V_{\cal R}\over {\cal U}_{mod,{\cal R}}} {1\over {(2\pi)^7}} \sqrt{{M\over K}}
\ee
In our regime of interest, we will have $K\gg M$ and $\beta\ll 1$, with the ratio ${V_{\cal R}\over{\cal
U}_{mod,{\cal R}}}\propto\beta^{1/3}$.  The expression (\ref{coreLtwo}) shows that we cannot decrease $\beta$
arbitrarily without causing the core size of our wrapped D4-brane to exceed $L_{u_2}$, requiring a new analysis
of the core size in the remaining $u_1,\tilde u_1$ directions.  (That is, the expressions (\ref{coreLtwo}),
(\ref{coreLone}) are only valid if both are small.)  However we will find that there is a substantial window in
which the ratios (\ref{coreLtwo}), (\ref{coreLone}) are highly suppressed.  This also implies that the core size
is much smaller than the curvature radii in our space and the probe approximation is valid (the curvature
invariants here, ${\cal N}_3$ being three dimensional, are $R,R^{mn}R_{mn},\det R/\det g$, and thus we get all
three curvature radii to be of the same order: $R^{-1/2},(R^{mn}R_{mn})^{-1/4},(\det R/\det g)^{-1/6}\sim
L_u^2/(M L_x)\gg L_{u_2}$).

\subsection{Basic Observational Constraints:  e-foldings and power spectrum}
\label{subsec:observational}

We must satisfy two basic requirements for the observational viability of the model -- conditions on the number
of e-foldings and the normalization of the scalar power spectrum.  At fixed $\beta$ these two conditions fix $M$
and $K$, so it will prove useful to retain the independent parameter $\beta$.  It is worth emphasizing that
because of the numerical factors in (\ref{sixform}), a large value of $K$ need not correspond to a large value
of the input flux quantum number $f_6$.

For any model with a single stage of inflation to be observationally viable, it must produce at least some 60
efolds of slow-roll inflation before the process exits to the minimum of the potential in order to solve the
isotropy, homogeneity and entropy problems of standard hot big bang cosmology~\cite{Inflation,Mukhanov:2005sc}.
For large field chaotic inflation with a power-law potential $V(\phi)\propto \phi^p$ the field $\phi$ has to
start at a value of \beq \phi_{N_e}=\sqrt{2pN_e}M_{\rm P} \eeq to generate $N_e$ efolds of slow-roll inflation
before inflation ends. The number of efolds is determined in slow-roll by \beq N_e=\int_{t_{N_e}}^{t_{\rm
exit}}H dt\simeq\int_{M_{\rm P}}^{\phi_{N_e}}\frac{d\phi}{M_{\rm
P}^2}\,\frac{V}{V'}=\frac{1}{2p}\,({\phi_{N_e}^2\over M_{\rm P}^2}-1)\quad. \eeq

Two basic observational conditions for the viability of our mechanism are
\noindent{(i)} Obtaining $\phi_{\rm max}$ larger than $\phi_{60}=2\sqrt{N_e/3}M_{\rm P}\simeq 9M_{\rm P}$ at
$N_e=60$ in our regime where $V_{\cal R}(\phi)=\mu^{10/3}\phi^{2/3}$, while also
\noindent{(ii)} obtaining the appropriate scale of the density perturbations at $N_e=60$ efolds before the end
of inflation.

This latter criterion is that we must generate a level of scalar curvature perturbation $\left.\Delta_{\cal
R}\right|_{60}\simeq 5.4\times 10^{-5}$. In our system,
\beq \left.\Delta_{\cal R}\right|_{N_e}=\left.\sqrt{\frac{1}{12\pi^2}\,\frac{V^3}{M_{\rm
P}^6V'^2}}\right|_{N_e}=\frac{(4/3)^{1/6}}{2\pi}\,N_e^{2/3}{\mu^{5/3}\over{M_{\rm P}^{5/3}}} \eeq
This depends on a combination of powers of $\beta$ and $L,L_x$ which is independent of the one appearing in
eq.~(\ref{phimax2}).  We will shortly write this in terms of the physical quantities $M$ and $K$ in the
construction \cite{ESdS}.  This will leave a candidate window for inflation -- consistent with moduli
stabilization, $N_e\simeq 60$, and $\left.\Delta_{\cal R}\right|_{60}\simeq 5.4\times 10^{-5}$ -- obtained with
only modest tuning of the parameters.  Then we will analyze the problem of controlling all ${\cal O}(10^{-2})$
contributions to $\epsilon$ and $\eta$ arising our string compactification.  Before turning to that, we pause to
also indicate the conditions for eternal inflation in our background, which would involve even greater
super-Planckian excursions in field space.

\subsection{Conditions for Eternal inflation}

The scalar curvature perturbations $\Delta_{\cal R}$ and in turn the primordial density perturbations are both
generated from the quantum fluctuations of scalar fields in a de Sitter background \beq
\sqrt{\langle\delta\phi^2\rangle_q}=\frac{H}{2\pi}\quad. \eeq They begin to dominate the classical slow-roll
motion once we have $\Delta_{\cal R}\gtrsim 1$. If this is case then inflation in that regime never ends but is
started over and again, rendering inflation in the global space-time eternal to the future. The boundary of
eternal inflation, $\phi_*$, can thus be determined from \beq \frac{1}{M_{\rm P}^3} \left.
\frac{V^{3/2}}{V'}\right|_{\phi_*}\gtrsim \sqrt{12\pi^2}\quad. \eeq

For a power law potential, the value of $\phi_*$ is thus fixed once the curvature perturbation at 60 efolds
before the end of inflation has been normalized to the COBE value: E.g. in the $\mu^{10/3}\phi^{2/3}$ under
study here the ratio $V/\epsilon$ is controlled by $\mu$ whose value is fixed by the COBE normalization to be
\beq \mu_{\rm obs.}\simeq 1.6\times 10^{-3}\quad. \eeq

Now $\phi_*$ is fixed once $\mu$ is observationally determined, while $\phi_{\rm max}$ can be adjusted to some
extent by decreasing $\beta$, though $\beta$ cannot be reduced arbitrarily without introducing problematic back
reaction (\ref{coreLtwo}) and complicating moduli stabilization (\ref{UKK}).  When it applies, eternal inflation
may considerably mitigate the problem of initial conditions for inflation.

\subsection{Specific results in a concrete Nil manifold construction of dS vacua}
\label{subsecspecific}

The results of the last sections concerning the appearance of a candidate large field chaotic inflation model
with a $\mu^{10/3}\phi^{2/3}$-potential for the D4-brane motion apply rather generally in compactifications of
type IIA string theory to 4d on Nil manifolds.

However, the parameters of the model depend on the moduli VEVs $L, L_x, L_u,$ and $\beta$, which are determined
in terms of topological, brane, and flux quantum numbers in any concrete compactification.  To assess the
viability of our mechanism in a concrete model, we therefore include in this section the formulas for our model
parameters in terms of these discrete quantum numbers in the construction \cite{ESdS}\ on a product ${\cal
N}_3\times{\cal \tilde N}_3$ of Nil three-manifolds.
It was shown there that the VEVs of the potentially runaway moduli in their dS minimum scale in terms of the
brane and flux quantum numbers $M,K$ as
\beq \label{cfactors} L=c_L\cdot K^{1/6}\quad,\quad L_x=\frac{c_{L_x}}{M^{1/2}}\quad,\quad g=\frac{\hat
g}{K}\quad. \eeq
$c_L,c_{L_x}$ here are numerical constants, and for a generic Nil manifold compactification will be of ${\cal
O}(1)$.  For the minimal setup described in \cite{ESdS}, we estimate their values as
\beq c_L\simeq (56\pi^2)^{1/12}\simeq 1.7\quad,\quad c_{L_x}=3^{1/8} 2^{5/4}\pi\simeq 8.6\quad,\quad \hat
g=\sqrt\frac{2}{7} (6+\sqrt 3)\pi^2\simeq 41 \quad. \eeq
We will include these specific numbers to get a feel for the magnitudes of the parameters that might arise in a
specific model realizing our mechanism -- however we emphasize that there are many variants of the construction
which will shift the values (\ref{cfactors}), and in the expression for our estimate of $\hat g$ we neglected
the KK5-brane contribution for simplicity (this would kick up $\hat g$ slightly).
In terms of $L^3=L_u^2L_x$ and $L_x$ we have then \beq L_u=\frac{c_L^{3/2}}{c_{L_x}^{1/2}}\,(KM)^{1/4}\simeq
0.75\times (KM)^{1/4}\quad. \eeq
We see that there is a controlled regime with $K\gg M>1$ for which $L, L_u\gg L_x$ and $L,L_u\gg 1$.  In
particular, we will be interested in a regime which satisfies $K\gg M$ numerically, without taking a limit where
$K/M$ becomes parametrically large; a sample numerical solution of this kind was described in \S3.8 of
\cite{ESdS}.  This setup is simplest in that it contains the fewest ingredients, but is subject to several
subtleties noted in \cite{ESdS}: the KK and moduli mass matrices are positive definite in themselves, but the
question of whether unstable directions might arise due to their mixing has not been analyzed in detail.

We can now plug these results into the general formulas above for the values $\phi_{\rm crit}$, $\phi_{\rm max}$
and $\phi_*$ (describing the boundaries to the $m^2\phi^2$-regime, the destabilization of the moduli, and the
regime of eternal inflation, respectively) as well as into the result for the curvature perturbation. We arrive
then at
\bea \frac{\phi_{\rm crit}}{M_{\rm P}}&\sim&(2\pi)^{3/2}\gamma^{-1/2}{{\hat g}^{1/2}\over\sqrt{2}}
\left(\frac{c_L}{c_{L_x}}\right)^{9/4}\nn\\
\frac{\phi_{\rm max}}{M_{\rm P}}&\sim&\frac{1}{3}\,\frac{\gamma}{\hat g (2\pi)^3}\left(\frac{c_L}{c_{L_x}}\right)^{-9/2}\\
\frac{\phi_*}{M_{\rm P}}&\sim& K^{9/8} \gamma^{1/4}\nn \eea and
\beq \left.\Delta_{\cal R}\right|_N\sim 60^{2/3} {(2\pi)^{7/2}\over{2^{5/6}}} K^{-3/2} \gamma^{-1/3}\hat g^{4/3}
c_L^{-1}\left(\frac{c_L}{c_{L_x}}\right)^{-1/2}\quad. \eeq
where we plugged in $N_e=60$. Here $\gamma$ denotes the combination
\be\label{gamdef}\gamma\equiv\beta^{-1/2}\left(\frac{K}{M}\right)^{1/4} \ee
This is larger than 1 in our regime of interest with $\beta\lesssim 1$ and $K>M$.

Now the ratio $c_L/c_{L_x}\simeq0.2$ in the concrete construction is consistent with our parametric result that
$\phi_{\rm crit}$ tends to be small (unless $\beta \gg1$), pushing us deep into the
$\mu^{10/3}\phi^{2/3}$-regime of these formulae, and that $\phi_{\rm max}\gg M_{\rm P}$ if $\beta\lesssim 1$ and
$K\gg M$.

More explicitly we can see that we need, firstly, $\gamma\gtrsim 190 $ for $\phi_{\rm max}>9M_{\rm P}$ to get at
least some 60 efolds of slow-roll inflation. Secondly, the observational constraint $\left.\Delta_{\cal
R}\right|_{60}=5.4\times10^{-5}$ implies $\gamma^{1/3} K^{3/2}\sim 1.9\times 10^{10}$, and putting this together
with the condition $\gamma\ge 190$ gives $K\le 2.2\times 10^6$, corresponding to a modest flux quantum number
$f_6\le 310$ (using (\ref{sixform})).   Putting this together with (\ref{gamdef}), we have that
\be \label{Mbeta}\beta M^{1/2}\le 0.04 \ee
Thus for $M\sim 1$, we obtain $\beta\sim 0.04$, and for $M\sim 10,\beta\sim .01$.

We note that the numbers we have obtained for $M$ and $f_6$ are close to those obtained in the sample numerical
solution discussed in the simplest version of the construction \cite{ESdS}\ (the version without additional
NS5-branes added to reduce the KK5-brane tensions).  This will be useful for us, since additional
moduli-stabilizing ingredients complicate the problem of suppressing contributions to the slow-roll parameters.
We will shortly comment on the open question of parametric limits with arbitrarily small couplings and
curvatures.

The number $M u_1$ of wrappings of our brane around the $x$ direction is now given by (\ref{windingN})
\be\label{windingNnumerical} N_w=Mu_1\sim {V_{\cal R}\over {{\cal U}_{mod,{\cal
R}}}}{1\over{(2\pi)^3\sqrt{2}}}{1\over{\hat g}}\left({c_{L_x}\over c_L}\right)^3(KM)^{1/2} \sim {V_{\cal R}\over
{{\cal U}_{mod,{\cal R}}}} 20 M^{1/2}\ee
and the ratio (\ref{coreLtwo}) of the core size $r_c$ to $L_{u_2}\sqrt{\alpha'}$ is consistently extremely
small.

A much stronger fine-tuning of $\beta$ (c.f. \cite{AccInf}) would be required to obtain eternal inflation, and
would need to be analyzed with respect to the back reaction criteria discussed above.

\subsubsection{Conditions for a Parametric Effect}
\label{subsubsec:parametric}

We note that so far in this section, we have fixed some of our parameters using the observed COBE normalization
for the scalar power spectrum.  At the resulting values of $K,M$, and $\beta$ we obtain results consistent with
the large-volume, weakly curved, locally supersymmetric weak-$g$ regime of the compactifications studied in
\cite{ESdS}\ (subject to the same subtleties with respect to separating the KK and moduli mass scales indicated
there).

At a theoretical level, one might wonder if the effect can be made ``parametrically large", increasing the field
range to be arbitrarily large while systematically improving the level of control so that the expansion
parameters become arbitrarily small (without necessarily imposing the condition $\left.\Delta_{\cal
R}\right|_{60}\simeq 5.4\times 10^{-5}$).

Let us comment on this question here.  Using the relations derived above, but not imposing the COBE
normalization on the power spectrum, we have the conditions
\be\label{paramrelns} \Delta_{{\cal R}}|_{\phi=\phi_{\rm max}}\sim 10^3 {\gamma\over K^{3/2}}\le 1 \quad\quad
{\phi_{\rm max}\over M_{\rm P}}\sim 0.1\gamma \to\infty \quad\quad {r_c\over{L_{u_2}\sqrt{\alpha'}}}\sim
{0.005\over{\beta^2\gamma^2}} \ll 1 \ee
where as above, $\gamma=\beta^{-1/2}(K/M)^{1/4}$.   From this we see firstly that in order to parametrically
increase the field range, we must take $\gamma$ large. Given that, in order to prevent arbitrarily large
curvature perturbations, we must increase $K$ such that $K^{3/2}\ge 10^3\gamma$.

As far as these formulae go, one could obtain such a parametric limit by taking $K$ sufficiently large (without
needing to adjust the anisotropy $\beta$).  This suppresses the coefficient $\mu^{10/3}$ in our potential
$\mu^{10/3}\phi^{2/3}$, extending the field range within the regime where the inflationary potential is smaller
than the moduli potential, and decreasing the back reaction parametrically.  In the construction \cite{ESdS},
increasing $K/M$ is in fact desirable also in order to increase the ratio of the lightest KK masses to the
heaviest moduli masses, and a method for achieving this was sketched in that work.  However, this method
requires a significant elaboration of the construction -- it involves introducing local sources of larger string
coupling, such as NS5-branes,  which intersect with the KK5-branes so as to lower their tension.  As we will see
in the next section, such defects generically complicate the problem of suppressing contributions to the
slow-roll parameter $\eta$, though a symmetric arrangement of them might make it possible to make use of this
more general construction to seek an explicit parametric limit of our large-field model.  Note that it is not
possible to parametrically increase the effect merely by decreasing $\beta$, since then the final condition in
(\ref{paramrelns}) -- that of avoiding large back reaction from the wrapped brane -- would eventually fail.

Regardless, again we note that the parameters we need for a self-consistent -- and observationally accurate --
model realizing our monodromy mechanism appear to be reasonable.  That is, they lie in a large-radius, locally
supersymmetric weakly curved, weak-$g$ regime of the Nil manifold compactification, suggesting that the model is
controllable as argued in \cite{ESdS}. Therefore we return to our main example, and turn to a systematic
analysis of the slow-roll parameters, and the corrections affecting them, in our setup.


\section{Theory Foregrounds:  Systematic Analysis of Corrections}
\label{sec:corrections}

So far we have seen that the curvature-induced potential $V_{\cal R}(\phi)$ provides a promising candidate
inflaton potential.  However, we must include all contributions to the full potential $V_{\rm
tot}(\phi,\phi_\perp)$ (as a function of $\phi$ and any other light scalar fields $\phi_\perp$ in the problem)
which affect the slow-roll parameters
\be \label{slowroll} \epsilon = {1\over 2}M_{\rm P}^2\left({V'\over V}\right)^2\lesssim 0.01 ~, ~~~~~~ \eta =
M_{\rm P}^2{V''\over V}\lesssim 0.01 \ee
at the ${\cal O}(0.01)$ level. The standard ``$\eta$ problem", for example, can be phrased (independently of the
scale of supersymmetry breaking) as the fact that a generic dimension six Planck-suppressed operator of the form
\be\label{badop} V(\phi){(\phi-\phi_0)^2\over M_{\rm P}^2} \ee
would make an ${\cal O}(1)$ contribution to $\eta$.

Moreover, we must ensure (\ref{slowroll}) holds at all points on the inflaton trajectory (ranging over
approximately 9 Planck units in field space in our $\phi^{2/3}$ model).  A priori, for a candidate large-field
model such as ours, this condition requires a functional fine tune of parameters in the effective action.  In
our setting, however, the monodromy of our D4-brane leads to the following simplification. Each time the
D4-brane moves around the $u_1$ direction, it becomes heavier as it wraps a longer cycle, but it encounters the
same background sources contained in the stabilized compactification. This means that as long as we take into
account the longer cycle wrapped by the brane in each interval $\Delta u_1=1$, the methods for analyzing
corrections to the potential are essentially the same in each such (sub-Planckian) interval.  Thus if
small-field brane inflation is controllable, similar methods may apply in the present setting to render
large-field inflation controllable.

\subsection{Contributions to $V,\epsilon$, and $\eta$ from Moduli-Stabilizing Ingredients}

\smallskip

Let us start by addressing the contributions to the background fields in (\ref{DfourSgen}) which arise from the
ingredients in the construction \cite{ESdS}\ which go beyond the curvature of the Nil manifold already
considered.  As discussed in \cite{KKLMMT,BDKMS,BDKM,Sequestering}, localized sources involved in moduli
stabilization often lead to order one contributions to slow-roll parameters, some of which are rather subtle.
For example, the interaction energy between the brane and other localized sources gets important contributions
both from modes propagating directly between them and from modes propagating from one to the other by going
around the compactification. In our problem, the homogeneity of the underlying Nil manifold combined with the
symmetries and the extended nature of the sources of moduli stabilization will help suppress these effects. Let
us start, however, by reviewing how this problem arises explicitly in our setup.

Let us begin by determining the effects of other localized defects (branes and orientifolds).  In the
construction \cite{ESdS}, the compactification manifold is a product ${\cal N}_3\times \tilde{\cal N}_3$ of two
Nil 3-manifolds of the form (\ref{nilgeom}) with coordinates $u_1,u_2,x$ and $\tilde u_1,\tilde u_2,\tilde x$
respectively. It is convenient to dimensionally reduce on the $x,\tilde x$ directions, since these are
stabilized at a small radius in the construction. Doing this, we can write schematically the general form of the
interaction potential energy $\Delta V_{4,X}$ between our D4-brane and other local sources $X$:
\be\label{deltaVgen} \Delta V_{4,X} = G_{N,8}\int d^4 u L_u^4(\alpha')^2 \int d^4 u'L_u^4(\alpha')^2{{\rho_4(u)
\rho_{X}(u')}\over{L_{u_1}^2\alpha'(\vec u_1-\vec u_1')^2 +L_{u_2}^2\alpha'(\vec u_2-\vec u_2')^2 }} \ee
where $G_{N,8}=\Gamma(7/2)2^5\pi^{7/2}(\alpha')^3g_s^2/L_x^2$ is the eight-dimensional Newton constant, and
$\vec u_i$ refers to the two-component vector $u_i,\tilde u_i$ for each $i=1,2$.  The energy densities
$\rho_4(u)$ and $\rho_X(u')$ are delta-function localized at the positions of the sources, and proportional to
the tension of the object.  For the D4-brane, $\rho_4(u)$ is proportional
$\tau_4={1\over{(2\pi)^4(\alpha')^{5/2}g_s}}$.

Let us start with the KK5-branes in the construction since they contribute potential energy at tree level. The
total KK5-brane tension is $n_K\tau_{KK5}=\zeta n_K{L_x^2\over{\beta^{1/2}(2\pi)^7g_s^2(\alpha')^3}}$ (where
$n_K$ is the number of KK5-branes, required to be a multiple of $M$ in \cite{ESdS}, and where $\zeta$ depends on
the local value $e^{\phi_{loc}}$ of the string coupling in the region where the KK5-brane sits \cite{ESdS} :
$\zeta=g_s^2e^{-2\phi_{loc}}$).

The interaction energy of the D4-brane with the KK5-branes depends on how they are oriented relative to each
other and on the distance between them.  After dimensionally reducing on the $x,\tilde x$ directions, each wraps
a one-cycle within the four directions $u_1,u_2,\tilde u_1,\tilde u_2$. We will consider the generic case that
the KK5-branes are not parallel to the D4-brane.

Let us analyze their interaction energy first in the case that they are widely separated. This will produce an
$\eta$ problem in the direction of their relative separation, as in \cite{KKLMMT}. For a large distance
$u_{4,KK5}L_u\sqrt{\alpha'}\gg \sqrt{\alpha'}$ between our D4-brane and the KK5-brane (here not keeping track of
the anisotropy $\beta$, which will enter in our more precise analysis below), we obtain
\be \label{VfourKK}\Delta V_{4,KK5}\sim {1\over {2^2\pi^{7/2}}}{\zeta n_K\over L_{u_2}}L_x V_{\cal R}(\phi) {\rm
log}({u_{4,KK5}}) \sim {\zeta\over {2^2\pi^{7/2}}}\left({M\over K}\right)^{1/4}V_{\cal R}(\phi) {\rm
log}({u_{4,KK5}}) ~~~~ ({\rm far ~field})\ee
where in the last step we indicated how this contribution scales parametrically with $M$ and $K$, in the
solutions \cite{ESdS}.  In that setup, the varying string coupling encoded in $\zeta$ is necessary in order to
obtain an arbitrarily small ratio $M/K$ :   $\zeta\lesssim {\cal O}((M/K)^{1/4})$ in this parametric limit.
However, for modest values of the parameters, such as the example studied numerically in \cite{ESdS}, and for
the minimally tuned model discussed above in \S\ref{subsecspecific}, we have that this ratio $M/K$ is
numerically small.

This contribution (\ref{VfourKK}) to the potential, and to the slow-roll parameters in the $\phi$ direction,
evidently becomes small in the regime $M\ll K$.  However, in this far field configuration, we must check whether
significant relative motion between the KK5-brane and the D4-brane is induced by the potential between them. If
that happened, the corresponding scalar field $\phi_{4,KK5}$ would either become the inflaton, or -- more
generically -- would roll too rapidly to produce inflation.

To check this, since $3 H\dot \phi=-\del_\phi V$ and $3 H\dot \phi_{4,KK5}=-\del_{\phi_{4,KK5}}V$, to ensure
that $\dot\phi_{4,KK5}\ll\dot\phi$, we must insist that
\be \label{singlefieldI} |\del_\phi V|\sim {{V_{\cal R}(\phi)}\over\phi}\gg |\del_{\phi_{4,KK5}}V|\sim
{\zeta\over{2^2\pi^{7/2}\phi_{4,KK5}}}\left({M\over K}\right)^{1/4}V_{\cal R}(\phi) \ee
which implies
\be \label{singlefieldII} {\zeta\over {2^2\pi^{7/2}}}\left({M\over K}\right)^{1/4}\ll{\phi_{4,KK5}\over\phi} \ee
Canonically normalizing the field $\phi_{4,KK5}$, we find for the motion mode $\phi_{4,KK5}$ of the KK5-brane
\be {\phi_{4,KK5}\over\phi}\sim {M_{\rm P}\over\phi}{\zeta^{1/2}\over (2\pi)^{7/2}}\left({M\over
K}\right)^{1/8}\ee
(In this discussion, we are assuming for now -- most conservatively -- that no other effect masses up this
mode.) For the realistic model, we have ${\phi\over M_{\rm P}}\sim 9$, and the condition (\ref{singlefieldII})
might be possible to solve, by arranging that $\zeta(K/M)^{1/4}$ be somewhat smaller than 1.

However, in order to avoid a large $\eta$ parameter in the $\phi_{4,KK5}$ direction, we would need to require --
taking two derivatives of $V$ -- that
\be \label{singlefieldIII} {\zeta\over {2^2\pi^{7/2}}}\left({M\over K}\right)^{1/4}\le
\left({\phi_{4,KK5}\over\phi}\right)^2 ~~~~ {\rm for} ~ \eta_{\phi_{4,KK5}}\le\eta_\phi \ee
This, however, does {\it not} hold parametrically:  the RHS of (\ref{singlefieldIII}) is in fact {\it smaller}
by a factor of $(M_{\rm P}/\phi)^2\ll 1$.  Therefore without compensating contributions to the mass, we would
have an $\eta$ problem in this configuration; the interaction energy between the D4-brane and the KK5-branes
produces an order 1 contribution to $\eta$. Depending on the precise relative orientation of our ingredients,
the curvature, a varying string coupling, and orientifold group actions can contribute to the mass term for
$\phi_{4,KK5}$, and can potentially be used to avoid or cancel against the offending contribution
(\ref{singlefieldIII}), as we now sketch.

\subsubsection{A specific setup for string-theoretic $\phi^{2/3}$ inflation}



\begin{figure}[t]
\begin{center}
\includegraphics[width=16cm]{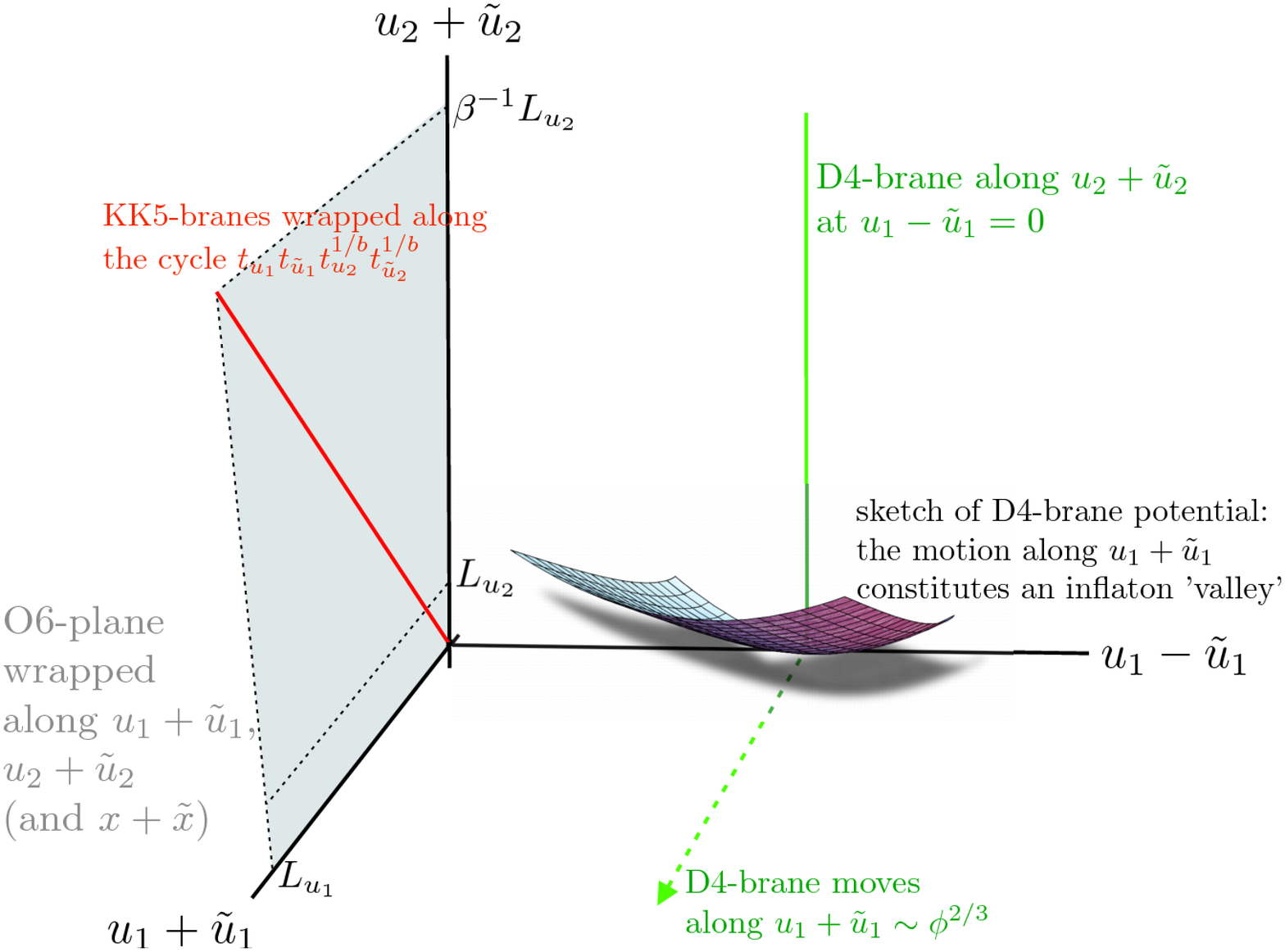}
\end{center}
\refstepcounter{figure}\label{branesetup3}

\vspace*{-.2cm} {\bf Figure~\ref{branesetup3}:}  The configuration of moduli-stabilizing ingredients and their
relation to our D4-brane, in the setup outlined in the text as a method for controlling the slow-roll
parameters.
\end{figure}

As a specific example (depicted in Figure \ref{branesetup3}), wrap the KK5-branes on the cycle generated by the
translations $t_{u_1} t_{{\tilde u}_1} t_{u_2}^{1/b} t_{{\tilde u}_2}^{1/b}$, with $b$ taken to be of order
$\beta$ so as to stabilize $L_2/L_1$ at the value $\beta$. Wrap the D4-brane on the cycle generated by
$t_{u_2}t_{{\tilde u}_2}$, and place it at the point $u_1-\tilde u_1=0$ in the $u_1-\tilde u_1$ direction. It is
not mutually BPS with respect to the O6 plane considered in \cite{ESdS}.  Our strategy will be to separate this
point from the location of the O6-plane, and use the local curvature-induced mass obstructing the D4-brane's
motion in the $u_1-\tilde u_1$ direction to metastabilize the D4-brane away from the O6-plane and the KK5-brane.
(Alternatively, one may try to keep the D4 and KK5 branes together, a scenario we discuss in the context of the
$\tilde\phi^{2/5}$ model in \S\ref{subsubsec:twofifthsKKfive}\ below.

To this end, take the orientifold action to be the standard worldsheet orientation reversal $\Omega (-1)^{F_L}$
combined with the action $u_2\to \tilde u_2$, $u_1\to\tilde u_1+1/2$, $x'\to\tilde x' -M\tilde u_2/2$.  This
puts the orientifold 6-plane, the fixed point locus of this action, at $u_1=\tilde u_1+1/2,u_2=\tilde u_2,
x'=\tilde x'-M\tilde u_2/2$.  On the covering space of the orientifold action, it maps the D4-brane to an
anti-D4-brane, so the D4-brane becomes unstable if it gets too close to the O6-plane in the orientifolded space.
This is easiest to see perhaps by formally T-dualizing our O6-D4 system to an O2-D0 system. That our D4-brane
carries no net charge in the orientifold can also be seen from the fact that its potential field
$C^{(5)}_{0123,u_2+\tilde u_2}$ is odd under the orientifold action by virtue of the intrinsic parity under
$\Omega (-1)^{F_L}$ carried by $C^{(5)}$. Place the KK5-brane at the same position $u_1-\tilde u_1 =1/2$ as the
O6-plane; the orientifold action freezes its motion mode in this direction, projecting out the corresponding
transverse motion mode $\phi_{KK5}$.
Inflation corresponds to motion of the D4-brane in the $u_1+\tilde u_1$ direction, i.e. along the direction $u$,
where $u_1={\tilde u}_1\equiv u$ (and at fixed $u_2-{\tilde u}_2$).

Consider now the curvature-induced potential for D4-brane motion in the $u_1-\tilde u_1$ direction.  Setting
$u_1=u+\Delta u$ and $\tilde u_1=u-\Delta u$, this contribution to the D4-brane potential energy is of order
\be\label{DeltaVDeltau}  V_{D4,{\cal R}}(u,\Delta u) \sim {L_xM\over{(2\pi)^4 g_s(\alpha')^2}}\sqrt{(u+\Delta
u)^2+(u-\Delta u)^2}\sim V_{\cal R}(\phi)\left(1+{1\over 2}{(\Delta u)^2\over u^2}+\dots\right)  \ee
where we used the full metric on ${\cal N}_3\times{\cal N}_3$, taking into account that the D4-brane is wrapped
along the coordinate locus $u_2=\tilde u_2$.

The KK5-brane and the O6-plane also contribute to the potential for $\Delta u$.  Note that they are arranged
symmetrically with respect to the candidate inflaton direction $u$, so do not contribute any $\phi$-dependent
corrections.  We must check whether the curvature-induced mass just computed in (\ref{DeltaVDeltau}) suffices to
stabilize the D4-brane against motion away from $\Delta u=0$, or whether instead there is a tadpole from the
KK5-D4 force which shifts D4 position by a separation $\Delta u$ which is bigger than order 1.

The D4-O6 force comes in the the open string channel from strings stretching between the D4-brane and its
anti-D4-brane partner on the covering space of the orientifold.  The KK5-brane, as a leading contribution to the
moduli potential, is heavier and should yield the strongest contribution to the potential in the $\Delta u$
direction. We can analyze this starting again from our expression for the interaction energy $\Delta V_{4,KK5}$
(\ref{VfourKK}), now including the appropriate factors of $\beta$ which descend from the KK5 tension and from
the fact that we separate the branes in the $u_1-\tilde u_1$ direction (whose proper length scales like
$L_{u_1}(u_1-\tilde u_1)$). This gives
\be\label{VfourKKDeltau} \Delta V_{4,KK5}\sim {\beta^{1/2}\over{2^2\pi^{7/2}}}{\zeta ML_x\over L_2}V_{\cal
R}\left(\Delta u-{1\over 2}(\Delta u)^2+\dots \right) \ee
Thus in order to suppress the shift $\Delta u$, as well as the negative contribution to the mass squared, coming
from the D4-KK5 interaction energy, we must require
\be\label{KKsupress} {\zeta \beta^{1/2} M L_x u^2\over{2^2\pi^{7/2}L_2}}\lesssim {\cal O}(1) \ee
The left hand side is $\lesssim\zeta\beta^{1/2}/M^{1/2}$ in our specific example discussed in
\S\ref{subsecspecific}, and hence this condition can be satisfied.

We also find that this orientation is locally stable against tilting of the D4-brane. Tilting them increases the
length and hence introduces a mass term; this also turns out to dominate over the attractive tadpole between the
separated segments of D4 and KK5 (estimated from (\ref{VfourKK})), leading to only a small tilting of the brane.


\subsubsection{A setup for the version with a $\tilde\phi^{2/5}$ potential}
\label{subsubsec:twofifthsKKfive}

Let us also briefly discuss a mechanism for controlling these corrections to $\epsilon$ and $\eta$ in the case
of the $\tilde\phi^{2/5}$ model discussed in \S\ref{subsubsec:variant}.  If we wrap the D4-brane in the
direction $u_2-\tilde u_2$, then it is mutually BPS with respect to the O6-plane, to leading order.  In this
case, we can deal with the mutual attraction between the D4-brane and the KK5-brane by placing them together so
that they intersect.  We can maintain this throughout the evolution if we move the D4-brane appropriately along
$u_2+\tilde u_2$ as well as $u_1+\tilde u_1$.  Motion in both these directions corresponds to a model with a
$\tilde\phi^{2/5}$ potential as outlined in \S\ref{subsubsec:variant}.  In this example, as in the previous one,
the O6-plane is also homogeneously extended along the direction of motion of the D4-brane.

In this configuration, however, the D4-brane carries charge since it is mutually BPS with respect to the
O6-plane. So let us discuss the effects of the $\overline{\rm D4}$-brane charge which must also be included in
this version of the construction. We could include an explicit $\overline{\rm D4}$-brane, separated from the
D4-brane, for example, and try to locally stabilize it or otherwise cancel its contribution to $\eta$ using
other effects.  However, a cleaner approach is to use fluxes to cancel the brane charge, as in~\cite{GKP}. In
our case, the Chern-Simons term $\int C_5\wedge F_2\wedge H$ yields a source for the RR potential coupling to
the D4-brane ($C_5$) from internal RR and NS fluxes $\int F_2\wedge H$.

In order to use this method, we need to ensure that the added RR 2-form flux is compatible with the rest of the
ingredients used for moduli stabilization.  First, note that as reviewed in \cite{ESdS}, the $Z_2$ symmetry by
which we orientifold to produce the O6-plane acts with a $(-1)$ on $C_1$ and $C_5$ as well as on $B_2$, in
addition to its geometrical action exchanging the tilded and un-tilded coordinates.  This is consistent with the
coupling we wish to use, given the orientation of our D4-brane along $u_2-\tilde u_2$:  it implies that
$C_5=C_{5,0123{u_{2-}}}(du_2-d{\tilde u}_2)\wedge dx^0\wedge dx^1\wedge dx^2\wedge dx^3$, which is altogether
invariant under the orientifold action. The $H$ flux in \cite{ESdS}\ (eqn 3.13) is also of course invariant
under the O6 action, so the introduction of an invariant 2-form RR flux of the form
$F_2=Q_2[du_1\wedge(dx'+Mu_1du_2)-d{\tilde u}_1\wedge (d{\tilde x}'+M{\tilde u}_1d{\tilde u}_2)]$ is compatible
with the symmetry by which we wish to orientifold.  We must also check that the potential energy ${\cal
U}_{F_2}$ introduced by $F_2$ is compatible with moduli stabilization.  This holds as well:
\be\label{Ftwoenergy} {\cal U}_{F_2}\sim M_{\rm P}^4{g_s^4\over L^{6}}{Q_2^2\over{L_x^2 L_{u_1}^2}}\sim
\beta\left({M\over K}\right)^{1/2} Q_2^2 ~{\cal U}_{mod, {\cal R}} ~~ \ll ~~ {\cal U}_{mod,{\cal R}} \ee

\subsubsection{General comments on the $\eta$ problem}

As emphasized above, in order to produce a reliable model of large-field inflation, our control over $\epsilon$
and $\eta$ must hold over the entire super-Planckian range of $\phi$. In the present top-down construction, we
use the methods just described to control the dynamics of the D4-brane in each $\Delta (u_1+\tilde u_1)=1$
interval. As the D4-brane moves around the $u_1+\tilde u_1$ direction multiple times, its potential energy
increases as it wraps a longer cycle on the $T^2$ traced out by $x'+\tilde x'$ and $u_2+\tilde u_2$.  But the
basic methods we outlined to control its potential apply in each such interval.  Thus it seems that roughly
speaking, the problem of controlling inflation in this setup -- even large-field inflation -- reduces to the
problem of computing and suppressing contributions to $\epsilon$ and $\eta$ for brane inflation \cite{KKLMMT}\
in a sub-Planckian range of field space.

Even given this, the problem of controlling inflation in each sub-Planckian range of field space is itself
extremely subtle, as explained in the series of works \cite{KKLMMT,BDKMS,BDKM,Sequestering}.  For example in
brane inflation, the interaction energy between the D-brane and other branes or localized moduli stabilizing
ingredients get significant $\phi$ and volume-dependent contributions from fields propagating around the compact
manifold (as well as from fields propagating directly between the defects) \cite{Sequestering,KKLMMT}.  This is
true as well in our case, but with the specifications we have made the symmetries cancel some of these
contributions against each other: the D4-brane's collective coordinate $u_1+\tilde u_1$ does not correspond to
the distance between the brane and other localized sources of stress-energy at leading order.\footnote{See e.g.
\cite{IT}\ for a different use of symmetries, within a small-field inflationary scenario, to cancel
contributions to the slow-roll parameters.}


We should reiterate another set of subtleties in this construction. We have been considering the simplest
version of the construction \cite{ESdS}, which leaves a degeneracy between the heaviest moduli and lightest KK
modes as discussed there.  Additional ingredients (such as NS5-branes) introduced to split these scales would
entail further potential contributions to the candidate inflaton trajectory.  As discussed in \cite{ESdS}, the
angular moduli are also rather subtle, but the moduli potential does seem to exhibit local minima in these
directions whose precise location we have not computed.  In any case, the mechanism we propose in the present
work would apply much more generally in compactifications in which branes undergo monodromies, assuming they can
be stabilized with ingredients manifesting the requisite symmetries.

All existing constructions have some subtleties of this sort, and it is fair to say that small-field models seem
a priori easier to implement via a tuned cancellation of corrections to the inflaton mass within a small range
of $\phi$. In some ways, however, the present case is somewhat more straightforward to analyze than, say, a
warped Calabi-Yau manifold, as the compactification manifold ${\cal N}_3\times \tilde{\cal N}_3$ itself is
extremely simple.  Of course more generally, manifolds with metric flux are more generic and perhaps typically
more complicated than Calabi-Yau manifolds.  It would be very interesting to understand the range of possible
behaviors of the potential in a wider class of manifolds with metric and generalized fluxes.

\subsubsection{The Standard Model}

We should emphasize that our construction does not yet include an explicit Standard Model sector.  This is of
course necessary for a fully realistic model, and may be challenging since each additional defect may lead to a
new source of large corrections to the slow-roll parameters.  One may approach this by considering e.g. a
``brane box" construction, with D4-branes suspended between 5-branes, or an intersecting D6-brane construction.
In each case, the effects on moduli stabilization and on the inflaton potential would need to be analyzed and
controlled. We leave this for future work. Including the Standard Model is also a prerequisite for a detailed
study of reheating (see \cite{Green:2007gs}\ for a recent analysis of this issue in N-flation).

\subsection{$\alpha'$ and Loop Corrections}

Let us next estimate the $\alpha'$ and loop corrections to $\epsilon$ and $\eta$.  The $\alpha'$ corrections
arise from corrections to $\Phi(X)$,  $G_{MN}(X)$, and fluxes to which the D4-brane couples through the
effective action (\ref{DfourSgen}), as well as corrections to the form of this action from higher derivatives of
the bulk fields. To get a sense of the issues, us start with corrections to the potential from $\alpha'$ effects
($V\to V+\Delta_{\alpha'}V$) which are schematically of the form
\be\label{genalpha} \Delta_{\alpha'} V\sim V_{\cal R}\sum_n c_n({\cal R}\alpha')^n \ee
(where ``$({\cal R}\alpha')^n$ refers to scalars made from appropriate contractions of components of the Riemann
and metric tensors) plus similar terms in which the curvature is replaced by fluxes or dilaton gradients, where
$c_n$ are order one constants. In our setting of 10d type II string theory, the $n=4$ term is the leading
$\alpha'$ correction to the bulk metric. Each term in (\ref{genalpha}) gives a correction to $\epsilon$ and
$\eta$ of order
\be \label{sralpha} \Delta_{\alpha'}~\epsilon \sim {\partial\over{\partial(\phi/M_{\rm P})}}({\cal R}\alpha')^n,
~~~~~~ \Delta_{\alpha'}~\eta \sim \epsilon \Delta\epsilon + {\partial^2\over{\partial(\phi/M_{\rm P})^2}}({\cal
R}\alpha')^n \ee

Since the curvatures (and other field strengths) and their derivatives
are small in our background, we find that the effect of the $\alpha'$ corrections on the slow variation
parameters is small.  In our regime $\phi\gg\phi_{\rm crit}\Rightarrow L_xM u_1\gg L_{u_2}$, all components of
the metric and Riemann tensors of the Nil manifolds are proportional to an order 1 power of $\phi$. (They all
depend on $u_1$, if at all, through a combination of the form $c_1 L_{u_2}^2+L_x^2M^2u_1^2$ where $c_1$ is of
order 1.) Therefore their derivatives are at most of order
\be\label{derivform} M_{\rm P}^n\del_\phi^n ({\cal R}\alpha')^{n'} \lesssim ({M_{\rm P}\over\phi})^n ({\cal
R}\alpha')^{n'} \ee
Similarly,
\be\label{derivformtwo} M_{\rm P}^n\del_\phi^n V_{\cal R} \sim ({M_{\rm P}\over\phi})^n V_{\cal R} \ee
so since ${M_{\rm P}\over\phi}\sim 10^{-1}$ and  the curvatures ${\cal R}\alpha'$ are small, the corrections to
$\epsilon$ and $\eta$ from terms of the form (\ref{genalpha}) are negligible in our background.

%
%
%
%


More precisely, we recall the derivation of the DBI-form of the D-brane action from the calculation of the beta
functions of the boundary action in the worldsheet sigma model~\cite{CallanDBI}. It was shown there by the
structure of worldsheet 1-loop diagrams determining the beta function of the coupling corresponding to the world
volume gauge field $A_M$ that the DBI-form of the D-brane action holds for all orders in $\alpha'$ up to first
derivatives of $\Phi, B_{MN}\ \& \ F_{MN}$ and second derivatives in $G_{MN}$ and through 1-loop in the string
coupling. To this order therefore all corrections thus appear through the corrected $G_{MN}, B_{MN}$, and $\Phi$
alone which has already been taken into account.  Let us now analyze the first curvature corrections which can
appear in this action at the level of second derivatives of $G_{MN}$.


Schematically, the leading correction at ${\cal O}(\alpha')$ in the curvature to the D4-brane action is by the
symmetries of the action
\beq S_{\rm D4}\sim\int d^5\xi\sqrt{\det(G_{MN}+\alpha' c_{\cal R} {\cal R}_{MN})\del_\alpha X^M\del_\beta
X^N}(1+\tilde c_{\cal R}\alpha'{\cal R}+\dots) \eeq
where $c_{\cal R},\tilde c_{\cal R}={\cal O}(1)$.  The Ricci scalar ${\cal R}$ is small and independent of
position inside the manifold. The relevant nonzero components of the Ricci tensor are (for simplicity
suppressing the $\beta$ dependence)
\be\label{Riccitensor}  \alpha' R_{u_1u_1}=-{{L_x^2M^2}\over{2 L_u^2}} ~~~~ \alpha'
R_{u_2u_2}=-{L_x^2M^2\over{2L_u^4}}(L_u^2-L_x^2M^2u_1^2) \ee
Comparing these to the metric components $G_{u_1u_1}$ and $G_{u_2u_2}$ we see they do not correct the shape of
the potential and are highly suppressed.  (Higher powers of curvature could change the shape of the potential at
some order, but are further suppressed.)


Next let us address loop corrections to our background.  The center of mass position of our D4-brane is governed
by an abelian worldvolume theory.  In a flat spacetime background, the DBI action for the probe does not receive
loop corrections, and the absence of a potential term is guaranteed in that case by translation invariance.  In
our setup, the only corrections to the potential therefore arise by virtue of the ambient curvature, and are
suppressed as in the previous discussion (as well as by factors of the weak string coupling).  Nonetheless it is
interesting to consider the structure of loop corrections in the four-dimensional effective theory.

The quantum effective potential will have some dependence on the position of our D4-brane, and hence on $\phi$.
As a specific example of a 1-loop effect which manifests a $\phi$-dependent correction to the $\phi^{2/3}$ form
of our potential, consider the self-interactions of the scalar field perturbations $\delta\phi$ implied by our
leading potential $V_{\cal R}\sim \mu^{10/3}\phi^{2/3}\equiv \mu^{10/3}(\phi_0+\delta\phi)$ (where
$\phi_0=\phi_0(t)$ denotes the background evolution of the inflaton).  We have
\be\label{coupphi} V_{\cal R}\sim \dots+\lambda_3\delta\phi^3+\lambda_4\delta\phi^4+\dots \ee
with $\lambda_3\sim \mu^{10/3}/\phi_0^{7/3}\sim 10^{-12}M_{\rm P}$ and $\lambda_4\sim (\mu/\phi_0)^{10/3} \sim
10^{-13}$.  These terms generate, via 1-loop diagrams, corrections to the inflaton mass squared of order $\Delta
m_\phi^2\sim {1\over{(2\pi)^4}}(\lambda_3^2+M_{SUSY}^2\lambda_4)$, where $M_{SUSY}$ is the SUSY breaking scale
which cuts off the loop in the second case.  The inflaton mass squared in our original potential, $m_\phi^2\sim
\del_\phi^2 V_{\cal R}$, is of order $10^{-4}H^2\sim 10^{-12}M_{\rm P}^2$, much greater than these loop
corrections.

More generally, consider the structure of the 1-loop vacuum amplitude.  It depends on $\phi$ through the
dependence on the brane position $u_1$ of the masses of KK and string modes which propagate in the loop.
Schematically, the partition function is of the form
\be\label{partitionftn} \Delta V_{1-loop}\sim {1\over(\alpha')^2}\int d^4 k\int {d\tau_2\over\tau_2} \left(
\sum_{m_B,n_B}e^{-(m_B^2+p_{n_B}^2+k^2)\alpha'\tau_2}-\sum_{m_F,n_F}e^{-(m_F^2+p_{n_F}^2+k^2)\alpha'\tau_2}\right)
\ee
where $k$ is the four-dimensional momentum.  Here $m_B$ and $m_F$ are the boson and fermion string masses and
$p_{n_B}^2$ and $p_{n_F}^2$ are the boson and fermion KK mode mass squareds.  Integrating over $k$ and expanding
this out, we obtain as a conservative estimate for the size of this loop correction
\be\label{partitionexp} \Delta V_{1-loop}\le {1\over (\alpha')^2}(M_{SUSY}^2\alpha') \sim M_{\rm
P}^4{(2\pi)^{14}\over 4}g^4(M_{SUSY}^2\alpha')\sim  g (M_{SUSY}^2\alpha') {\cal U}_{mod,{\cal R}} \ee
where $M_{SUSY}$ is again the effective SUSY breaking scale which cuts off the loop, and where we used the fact
that all the leading terms in the moduli potential in \cite{ESdS}, such as the orientifold term $\sim
-g^3M_P^4$, scale like ${\cal U}_{mod,{\cal R}}$ in the solution. The leading effects which break both
supersymmetry and translation invariance in the $u_1$ direction come from the Nil manifold curvature, so we have
that $M_{SUSY}^2\alpha'$ is of the form
\be\label{SUSYcurv} M_{SUSY}^2\alpha'\sim \sum_n c_n({\cal R}\alpha')^n  \ee
Now since $g (M_{SUSY}^2\alpha') {\cal U}_{mod,{\cal R}}\ll {\cal U}_{mod,{\cal R}}$, there is a large window in
which $g (M_{SUSY}^2\alpha') {\cal U}_{mod,{\cal R}}\ll V_{\cal R} \le {\cal U}_{mod,{\cal R}}$.  Combining this
with the fact discussed above that derivatives with respect to $\phi/M_{\rm P}$ pull down inverse powers of
$\phi/M_{\rm P}$, we see that $M_{\rm P}^n\del_\phi^n \Delta V_{1-loop} \ll M_{\rm P}^n\del_\phi^n V_{\cal R}$,
which implies that the corrections to $\epsilon$ and $\eta$ from the 1-loop effective potential are small.

%
%

%
%

Let us next check the renormalization of Newton's constant as in \cite{Nflation}, given by
\be\label{Mprenorm} M_{{\rm P},renorm}^2=M_{\rm P}^2+N_s\Lambda^2 \ee
where $N_s$ is the number of species and where $\Lambda$ is the effective UV cutoff in loop momentum.. In
discussing the absence of back reaction of the D4-brane on the geometry in \S\ref{subsubsec:backreaction}, we
found the effective number of species added by our mechanism (\ref{windingN}).  In the specific example
discussed above in \S\ref{subsecspecific}, this was of order 20, or of order $10^2$ if we count the relatively
heavy open strings between the multiple wrappings of the brane.
%
%
Even if $\Lambda$ were as high as string scale, the correction is negligible since ${1\over\alpha'}\sim
{(2\pi)^7\over 2}{g^2} M_{\rm P}^2\sim {(2\pi)^7\over 2}{\hat g^2\over K^2}M_{\rm P}^2\sim 7\times 10^{-5}M_{\rm
P}^2$.

\section{Observational Predictions}

We shall thus now derive the observational predictions from this inflationary regime.  For the minimal case of a
single-field slow-roll model, our theory predicts negligible non-Gaussianity.  As such, it is cleanly
distinguishable from single-field models with low sound speed, and from a large class of multifield models.
(However, it is possible that the same mechanism could be generalized to cases with light transverse fields to
match an observed $f_{NL}^{local}$ if present.)

The remaining CMB observables are the spectral index $n_s$ of the curvature perturbation and the
tensor-to-scalar ratio $r$ of primordial gravitational waves generated during inflation. In the slow-roll
approximation these quantities at $N$ efolds before the end of inflation are given by \bea
\left.n_s\right|_N&=&1-6\epsilon_N+2\eta_N\\
\left.r\right|_N&=&16\epsilon_N \eea
where the index $N$ indicates that these quantities are to be calculated at about $N$ efolds before the end of
inflation, i.e. at $\phi_N$.

\begin{figure}[t]
\begin{center}
\includegraphics[width=15.5cm]{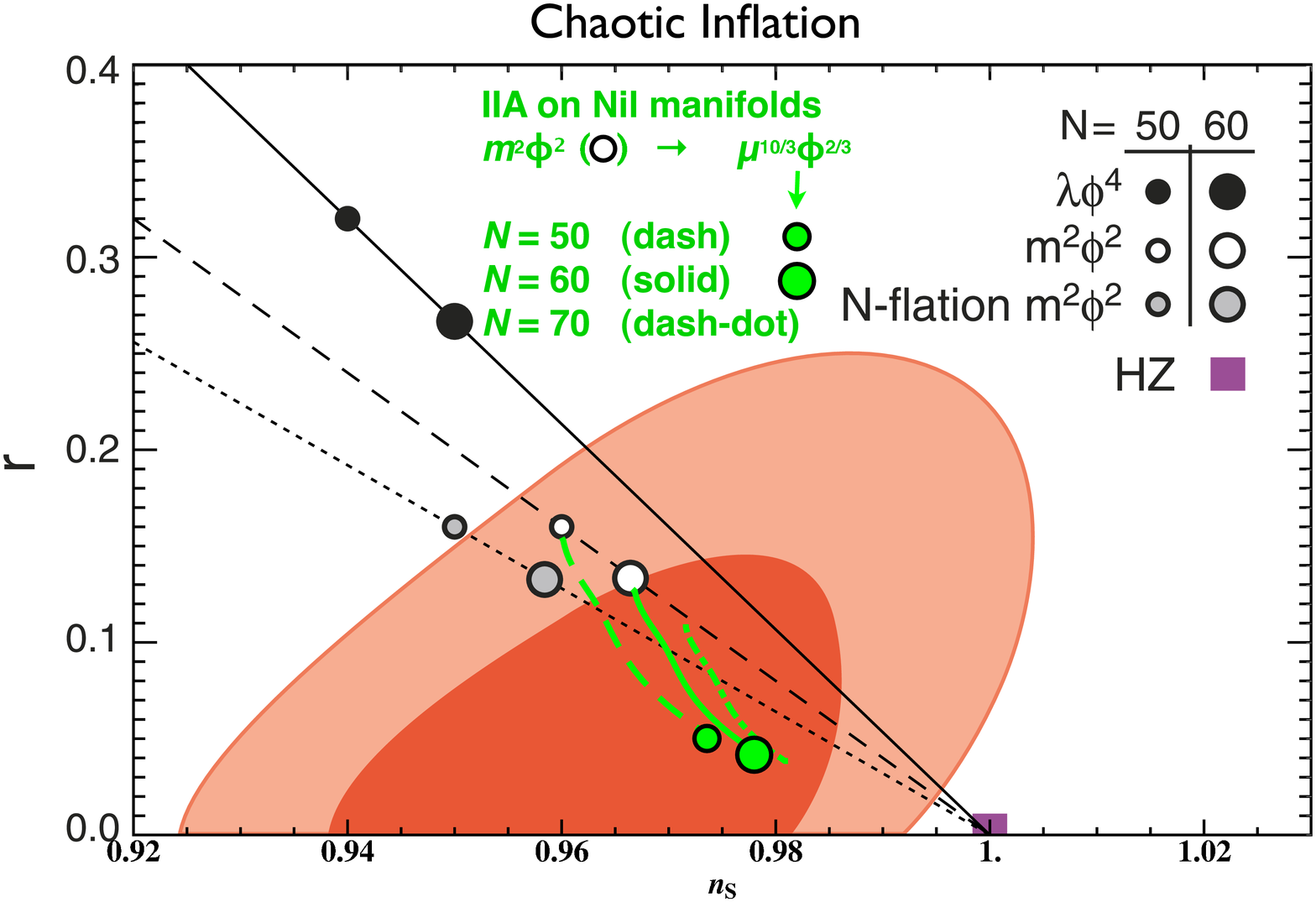}
\end{center}
\refstepcounter{figure}\label{fullpredict}

\vspace*{-.2cm} {\bf Figure~\ref{fullpredict}:} Red: 5-year WMAP+BAO+SN~\cite{Observations} combined joint 68 \%
and 95 \% error contours on $(n_s, r)$. Green: General prediction of the potential $V_{\cal R}(\phi)$
(\ref{pot1}) as one formally varies $\beta$ to interpolate between $m^2\phi^2$ (black hollow circles) and
$\mu^{10/3}\phi^{2/3}$ (green solid circles). Only the latter regime is viable in our setup as discussed in the
text, so the solid green circles (for $N=50,60$ efolds before the end of inflation) denote our prediction.
\end{figure}

Here, we can determine $n_s$ and $r$ in general only numerically as $V_{\cal R}(\phi)$ is given in its full form
only numerically. However, in the limiting case of a pure power law potential $V(\phi)\propto\phi^p$, these
observables are given by
\bea
\left.n_s\right|_N&=&1-\frac{2+p}{2N}\\
\left.r\right|_N&=&\frac{4p}{N}\quad. \eea

At about 60 efolds before the end of inflation, when the COBE normalization scale left the horizon, this yields
for $m^2\phi^2$ inflation $n_s=0.967$ and $r=0.13$.  For our case of the $\mu^{10/3}\phi^{2/3}$-limit of
$V_{\cal R}(\phi)$ we get
\beq n_s\simeq 0.978\quad{\rm and}\quad r\simeq 0.04\quad.\eeq
plus in general corrections which are at most of order 0.01 from e.g. the shifts in the moduli induced by the
inflaton potential (\ref{shifteta}).

Before taking into account moduli stabilization, the value of $\beta$ formally determines in which of these
regimes of the full potential the last 60 efolds of inflation fall, so we can numerically derive $r$ as a
function of $n_s$ parametrized by their dependence on $\beta$. This prediction is shown by the green curve in
Fig.~\ref{fullpredict} together with the 68\% and 95\% joint error contours of the 5-year WMAP
data~\cite{Observations} in the $(n_s,r)$-plane.

We see that the $m^2\phi^2$-endpoint (the upper left end of the green curve) matches with the open circle
denoting the pure $m^2\phi^2$-potential. Of course, the $m^2\phi^2$-regime in the string construction here is
not viable for the reason explained above that it destabilizes the moduli if we require the necessary 60 efolds
of slow-roll inflation. Thus, the observationally viable part of the green line consists of its lower right part
whose endpoint (the solid green circle) is to good approximation the pure $\mu^{10/3}\phi^{2/3}$-potential.

For the $\tilde\mu^{18/5}\tilde\phi^{2/5}$-case discussed above in \S\ref{subsubsec:variant}\ and
\S\ref{subsubsec:twofifthsKKfive}, we note that $n_s\simeq 0.98$ and $r\simeq 0.03$.

\section{Discussion}

Let us summarize what we have obtained.  By using a monodromy of wrapped branes on Nil manifolds -- the fact
that their approximate moduli space lives on the covering space of the compact twisted torus -- we extended the
kinematical field range of brane inflation in a simple way.  The resulting candidate inflaton has an
asymptotically power-law potential determined by the compactification geometry.  This potential is proportional
to $\phi^{2/3}$ in the case studied in detail, and we also find a variant with a $\tilde\phi^{2/5}$ potential
corresponding to a different direction of motion of the brane.

We analyzed a host of dynamical and observational conditions for a viable model. We imposed the condition that
the inflaton potential $V_{\cal R}(\phi)$ be subdominant to the moduli-stabilizing potential and cause
negligible back reaction in the geometry, and that the small $\phi$-dependent shifts in moduli not destabilize
the inflaton trajectory. These dynamical requirements proved to be consistent with motion over a super-Planckian
range of field space. We argued that by orienting our D4-brane in a symmetric manner with respect to the basic
moduli-stabilizing ingredients, we can avoid order 1 corrections to the slow-roll parameters $\epsilon$ and
$\eta$.  In analyzing this, we noted that the monodromy also provides some simplification of the problem that
one would a priori expect to need a functional fine-tune of the potential:  in our mechanism this large range in
field space is built up out of a set of similar, shorter segments corresponding to the motion of the brane once
around the compact manifold.

Putting these conditions together, we found that introducing a few  percent fine-tune (encoded in our parameter
$\beta$ and the flux quantum number $f_6$), we obtain a viable model realizing this mechanism.  The resulting
predictions for the tilt and tensor spectrum sit comfortably within the 1-sigma contours obtained from present
data, and are testable in upcoming experiments.

As explained in for example \cite{KL}, there is an interesting tension between the observation of gravity waves
and the scale of the moduli potential barrier (which can in some cases be related to the scale of supersymmetry
breaking).  Here we find that compactification manifolds with a larger potential barrier coming from negative
scalar curvature provide a reasonably natural mechanism for high-scale inflation.

We note again that our construction so far contains no Standard Model sector, whose inclusion would yield new
challenges for controlling the slow-roll parameters.  Moreover our underlying de Sitter compactification, like
all such constructions, has a number of subtleties (described in \cite{ESdS}).  In particular, the off-diagonal
elements in the mass matrix mixing $g,L,L_x$ with the angular moduli and with the lightest KK modes has not been
explicitly calculated, and depend on order 1 coefficients. There appear to be sufficient ingredients available
to tune these entries if necessary, but this has not been carried out explicitly. We leave a more detailed
analysis of these issues for future work.  Similar open questions arise for all string inflationary models, and
it would be interesting to work towards a fully explicit construction of inflation combined with moduli
stabilization.

Monodromies of the sort used here arise in a broad class of string compactifications, including more general
twisted tori and also non-geometrical spaces \cite{Albionmonodromy,Monodrofold}, as well as in motion on the
closed string moduli space in Calabi-Yau compactifications \cite{CYmonodromies}. It will be interesting to
characterize the potentials that arise in viable models much more generally, in the special cases where the
inflaton direction is sufficiently homogeneous to allow control over $\epsilon$ and $\eta$.

Let us also note that we chose to include a single wrapped D4-brane in the compactification, with potential
minimized at $u_1=0$, but another mechanism for inflation -- ``trapped inflation" \cite{KLLMMS}\ -- suggests
itself in this geometry. Since the $T^2$ traced out by $x',u_2$ is equivalent at each position $u_1=j/M$ (for
integer $j$), there is a place for a locally stabilized wrapped D4-brane at each of these places. If present,
such extra D4-branes introduce points with extra light species along the inflaton trajectory.  The inflaton
dumps some energy into these species, since they get produced when it hits these points.  This motivates a
careful study of the conditions for and predictions of trapped inflation \cite{WIP}.

In general, one may wish to assess how contrived a given construction of observationally testable inflation
looks from a model-building point of view.  Indeed, without considering specific mechanisms, the problem of
testing UV-sensitive contributions to inflation is very difficult a priori \cite{Generalshort}. The present
construction involves modest fine-tuning and a simple mechanism for enhancing the geometric field range (which
should arise in a wide class of compactifications), but requires symmetries used to suppress destabilizing
corrections to the slow-roll parameters (which entails special choices, as with all symmetry principles). It is
perhaps worth noting that this work began not as a direct attempt to engineer a gravitational wave signature
from string inflation, but simply as an investigation of the prospects for inflation of any kind in the setup
\cite{ESdS}; the present mechanism and \cite{WIP}\ are what emerged thus far from this study. Similar comments
apply to other corners of field theory and string theory with predictive inflationary signatures such as
\cite{AccInf,DBISky,Cosmicstrings,Curvature}.  The development of a detailed understanding of UV complete
cosmological solutions is just beginning, and it is not yet clear how to put a measure on the space of
inflationary mechanisms.  However, it is already apparent that upcoming data will be very effective in deciding
among wildly different possibilities for field- and string-theoretic inflationary mechanisms.

\section*{Acknowledgments}
We dedicate this work to A. Linde on the occasion of his 60th birthday celebration. We thank T. Banks, J.R.
Bond, G. Efstathiou, S. Kachru, R. Kallosh, C.-L. Kuo, A. Lawrence, A. Linde, J. Maldacena, L. McAllister, B.
Netterfield, L. Senatore, and D. Tong for useful discussions. The research of E.S. is supported by NSF grant
PHY-0244728, by the DOE under contract DE-AC03-76SF00515, and by BSF and FQXi grants. The research of A.W. is
supported in part by the Alexander-von-Humboldt foundation, as well as by NSF grant PHY-0244728.

\begingroup\raggedright\endgroup

\end{document}